\pdfoutput=1
\documentclass[12pt,a4paper]{article}
\usepackage[utf8]{inputenc}
\usepackage{ifthen} 
\newboolean{pdflatex}
\setboolean{pdflatex}{true} 

\newboolean{articletitles}
\setboolean{articletitles}{true} 

\newboolean{uprightparticles}
\setboolean{uprightparticles}{false} 

\def\paperauthors{LHCb collaboration} 
\def\paperasciititle{Measurement of the top-quark production cross-section and charge asymmetry at LHCb} 
\def\papertitle{Measurement of the top-quark production cross-sections and charge asymmetry at \lhcb} 
\def\paperkeywords{{High Energy Physics}, {LHCb}} 
\def\papercopyright{\the\year\ CERN for the benefit of the LHCb collaboration} 
\def\paperlicence{CC BY 4.0 licence}
\def\paperlicenceurl{https://creativecommons.org/licenses/by/4.0/}

\newif\ifEnableSectionTOCLinks
\EnableSectionTOCLinksfalse 

\usepackage[top=1in, bottom=1.25in, left=1in, right=1in]{geometry}

\usepackage{microtype}
\usepackage{lineno}  
\usepackage{xspace} 
\usepackage{caption} 

\usepackage{graphicx}  
\usepackage{color}
\usepackage{colortbl}
\graphicspath{{./figs/}} 

\usepackage{amsmath} 
\usepackage{amssymb}
\usepackage{amsfonts}
\usepackage{upgreek} 

\newcommand*\patchAmsMathEnvironmentForLineno[1]{%
\expandafter\let\csname old#1\expandafter\endcsname\csname #1\endcsname
\expandafter\let\csname oldend#1\expandafter\endcsname\csname
end#1\endcsname
 \renewenvironment{#1}%
   {\linenomath\csname old#1\endcsname}%
   {\csname oldend#1\endcsname\endlinenomath}%
}
\newcommand*\patchBothAmsMathEnvironmentsForLineno[1]{%
  \patchAmsMathEnvironmentForLineno{#1}%
  \patchAmsMathEnvironmentForLineno{#1*}%
}
\AtBeginDocument{%
\patchBothAmsMathEnvironmentsForLineno{equation}%
\patchBothAmsMathEnvironmentsForLineno{align}%
\patchBothAmsMathEnvironmentsForLineno{flalign}%
\patchBothAmsMathEnvironmentsForLineno{alignat}%
\patchBothAmsMathEnvironmentsForLineno{gather}%
\patchBothAmsMathEnvironmentsForLineno{multline}%
\patchBothAmsMathEnvironmentsForLineno{eqnarray}%
}

\usepackage[colorinlistoftodos,textsize=scriptsize]{todonotes}

\usepackage[bottom,flushmargin,hang,multiple]{footmisc}

\usepackage{xspace} 
\usepackage{upgreek}

\def\lhcb   {\mbox{LHCb}\xspace}
\def\atlas  {\mbox{ATLAS}\xspace}
\def\cms    {\mbox{CMS}\xspace}

\def\cdf    {\mbox{CDF}\xspace}
\def\dzero  {\mbox{D0}\xspace}

\def\lhc    {\mbox{LHC}\xspace}

\def\tevatron {Tevatron\xspace}

\def\MagUp {\mbox{\em Mag\kern -0.05em Up}\xspace}

\ifthenelse{\boolean{uprightparticles}}%
{

 \def\Pmu         {\ensuremath{\upmu}\xspace}

 \def\PDelta      {\ensuremath{\Delta}\xspace}                 
 \def\PXi         {\ensuremath{\Xi}\xspace}                 
 \def\PLambda     {\ensuremath{\Lambda}\xspace}                 
 \def\PSigma      {\ensuremath{\Sigma}\xspace}                 
 \def\POmega      {\ensuremath{\Omega}\xspace}                 
 \def\PUpsilon    {\ensuremath{\Upsilon}\xspace}
 \let\oldPi\Pi
 \def\PPi         {\ensuremath{\oldPi}\xspace}

 \def\PB      {\ensuremath{\mathrm{B}}\xspace}                 
 \def\PD      {\ensuremath{\mathrm{D}}\xspace}                 

 \def\PK      {\ensuremath{\mathrm{K}}\xspace}                 
 \def\PW      {\ensuremath{\mathrm{W}}\xspace}                 
 \def\PZ      {\ensuremath{\mathrm{Z}}\xspace}                 
 \def\Pb      {\ensuremath{\mathrm{b}}\xspace}                 
 \def\Pc      {\ensuremath{\mathrm{c}}\xspace}                 
 \def\Pd      {\ensuremath{\mathrm{d}}\xspace}                 

 \def\Pp      {\ensuremath{\mathrm{p}}\xspace}                 

 \def\Ps      {\ensuremath{\mathrm{s}}\xspace}                 
 \def\Pt      {\ensuremath{\mathrm{t}}\xspace}                 
 \def\Pu      {\ensuremath{\mathrm{u}}\xspace}                 
 \def\thebaroffset{0.0em}
}
{

 \def\Pmu         {\ensuremath{\mu}\xspace}

 \mathchardef\PDelta="7101
 \mathchardef\PXi="7104
 \mathchardef\PLambda="7103
 \mathchardef\PSigma="7106
 \mathchardef\POmega="710A
 \mathchardef\PUpsilon="7107
 \mathchardef\PPi="7105
 \def\PB      {\ensuremath{B}\xspace}                 
 \def\PD      {\ensuremath{D}\xspace}                 

 \def\PK      {\ensuremath{K}\xspace}                 
 \def\PW      {\ensuremath{W}\xspace}                 
 \def\PZ      {\ensuremath{Z}\xspace}                 
 \def\Pb      {\ensuremath{b}\xspace}                 
 \def\Pc      {\ensuremath{c}\xspace}                 
 \def\Pd      {\ensuremath{d}\xspace}                 

 \def\Pp      {\ensuremath{p}\xspace}                 

 \def\Ps      {\ensuremath{s}\xspace}                 
 \def\Pt      {\ensuremath{t}\xspace}                 
 \def\Pu      {\ensuremath{u}\xspace}                 
 \def\thebaroffset{0.18em}
}
\newcommand{\offsetoverline}[2][\thebaroffset]{\kern #1\overline{\kern -#1 #2}}%

\makeatletter
\ifcase \@ptsize \relax
  \newcommand{\miniscule}{\@setfontsize\miniscule{4}{5}}
\or
  \newcommand{\miniscule}{\@setfontsize\miniscule{5}{6}}
\or
  \newcommand{\miniscule}{\@setfontsize\miniscule{5}{6}}
\fi
\makeatother

\DeclareRobustCommand{\optbar}[1]{\shortstack{{\miniscule (\rule[.5ex]{1.25em}{.18mm})}
  \\ [-.7ex] $#1$}}

\def\mumu       {{\ensuremath{\Pmu^+\Pmu^-}}\xspace}

\def\W      {{\ensuremath{\PW}}\xspace}

\def\Z      {{\ensuremath{\PZ}}\xspace}

\def\uquark    {{\ensuremath{\Pu}}\xspace}

\def\dquark    {{\ensuremath{\Pd}}\xspace}

\def\squark    {{\ensuremath{\Ps}}\xspace}

\def\cquark    {{\ensuremath{\Pc}}\xspace}
\def\cquarkbar {{\ensuremath{\overline \cquark}}\xspace}
\def\ccbar     {{\ensuremath{\cquark\cquarkbar}}\xspace}
\def\bquark    {{\ensuremath{\Pb}}\xspace}
\def\bquarkbar {{\ensuremath{\overline \bquark}}\xspace}
\def\bbbar     {{\ensuremath{\bquark\bquarkbar}}\xspace}
\def\tquark    {{\ensuremath{\Pt}}\xspace}
\def\tquarkbar {{\ensuremath{\overline \tquark}}\xspace}
\def\ttbar     {{\ensuremath{\tquark\tquarkbar}}\xspace}

\def\KorKbar {\kern \thebaroffset\optbar{\kern -\thebaroffset \PK}{}\xspace}

\def\D       {{\ensuremath{\PD}}\xspace}

\def\DorDbar {\kern \thebaroffset\optbar{\kern -\thebaroffset \PD}\xspace}

\def\Dp      {{\ensuremath{\D^+}}\xspace}
\def\Dm      {{\ensuremath{\D^-}}\xspace}

\def\DpDm    {\ensuremath{\Dp {\kern -0.16em \Dm}}\xspace}

\def\B       {{\ensuremath{\PB}}\xspace}

\def\BorBbar {\kern \thebaroffset\optbar{\kern -\thebaroffset \PB}\xspace}

\def\Bd      {{\ensuremath{\B^0}}\xspace}

\def\BdorBdbar {\kern \thebaroffset\optbar{\kern -\thebaroffset \Bd}\xspace}

\def\Bs      {{\ensuremath{\B^0_\squark}}\xspace}

\def\BsorBsbar {\kern \thebaroffset\optbar{\kern -\thebaroffset \Bs}\xspace}

\def\Y#1S{\ensuremath{\PUpsilon{(#1S)}}\xspace}

\def\proton      {{\ensuremath{\Pp}}\xspace}

\def\LorLbar     {\kern \thebaroffset\optbar{\kern -\thebaroffset \PLambda}\xspace}

\newcommand{\decay}[2]{\mbox{\ensuremath{#1\!\to #2}}\xspace} 

\def\to                 {\ensuremath{\rightarrow}\xspace}

\def\AT#1     {\ensuremath{A_{\mathrm{T}}^{#1}}\xspace}           

\def\C#1      {\ensuremath{\mathcal{C}_{#1}}\xspace}                       
\def\Cp#1     {\ensuremath{\mathcal{C}_{#1}^{'}}\xspace}                    
\def\Ceff#1   {\ensuremath{\mathcal{C}_{#1}^{\mathrm{(eff)}}}\xspace}        
\def\Cpeff#1  {\ensuremath{\mathcal{C}_{#1}^{'\mathrm{(eff)}}}\xspace}       
\def\Ope#1    {\ensuremath{\mathcal{O}_{#1}}\xspace}                       
\def\Opep#1   {\ensuremath{\mathcal{O}_{#1}^{'}}\xspace}                    

\newcommand{\nospaceunit}[1]{\ensuremath{\text{#1}}}       
\newcommand{\aunit}[1]{\ensuremath{\text{\,#1}}}       

\newcommand{\tev}{\aunit{Te\kern -0.1em V}\xspace}
\newcommand{\gev}{\aunit{Ge\kern -0.1em V}\xspace}
\newcommand{\mev}{\aunit{Me\kern -0.1em V}\xspace}
\newcommand{\kev}{\aunit{ke\kern -0.1em V}\xspace}
\newcommand{\ev}{\aunit{e\kern -0.1em V}\xspace}
 
\newcommand{\mevc}{\ensuremath{\aunit{Me\kern -0.1em V\!/}c}\xspace}
\newcommand{\gevc}{\ensuremath{\aunit{Ge\kern -0.1em V\!/}c}\xspace}
\newcommand{\mevcc}{\ensuremath{\aunit{Me\kern -0.1em V\!/}c^2}\xspace}
\newcommand{\gevcc}{\ensuremath{\aunit{Ge\kern -0.1em V\!/}c^2}\xspace}

\def\mm   {\aunit{mm}\xspace}

\def\mum  {\ensuremath{\,\upmu\nospaceunit{m}}\xspace}

\def\pb {\aunit{pb}\xspace}

\def\fb   {\ensuremath{\aunit{fb}}\xspace}
\def\invfb   {\ensuremath{\fb^{-1}}\xspace}

\newcommand{\stat}{\aunit{(stat)}\xspace}
\newcommand{\syst}{\aunit{(syst)}\xspace}

\def\gsim{{~\raise.15em\hbox{$>$}\kern-.85em
          \lower.35em\hbox{$\sim$}~}\xspace}
\def\lsim{{~\raise.15em\hbox{$<$}\kern-.85em
          \lower.35em\hbox{$\sim$}~}\xspace}

\def\sqs   {\ensuremath{\protect\sqrt{s}}\xspace}

\def\pt         {\ensuremath{p_{\mathrm{T}}}\xspace}

\def\ptot       {\ensuremath{p}\xspace}

\newcommand{\lum} {\ensuremath{\mathcal{L}}\xspace}

\def\geant      {\mbox{\textsc{Geant4}}\xspace}

\def\photos     {\mbox{\textsc{Photos}}\xspace}
\def\powheg     {\mbox{\textsc{Powheg}}\xspace}
\def\pythia     {\mbox{\textsc{Pythia}}\xspace}

\def\tell1  {TELL1\xspace}
\def\ukl1   {UKL1\xspace}

\newcommand{\lhcborcid}[1]{\href{https://orcid.org/#1}{\hspace*{0.1em}\raisebox{-0.45ex}{\includegraphics[width=1em]{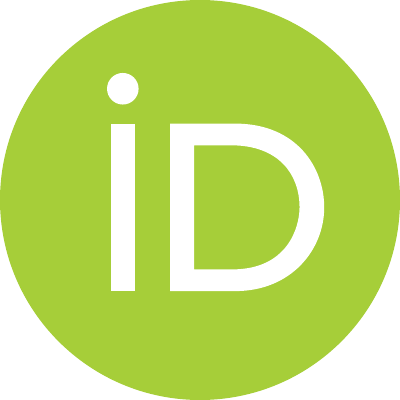}}}}

\def\Zmm     {\decay{\Z}{\mumu}}

\def\imu     {I_{\mu}\xspace}

\def\jet   {\text{jet}\xspace}

\def\bjet   {\ensuremath{\bquark\mathrm{\text{-jet}}}\xspace}

\ifEnableSectionTOCLinks
    \usepackage[explicit]{titlesec} 
    
    \let\oldcontentsline\contentsline
    \renewcommand

    \titleformat{\section}{\normalfont\Large\bf}{\hyperlink{tocsection.\thesection}{{\thesection} \parbox[t]{\dimexpr\textwidth-1pc}{#1}}}{1pc}{}

    \titleformat{\subsection}{\normalfont\bf}{\hyperlink{tocsubsection.\thesubsection}{{\thesubsection} \parbox[t]{\dimexpr\textwidth-1pc}{#1}}}{1pc}{}

    \titleformat{name=\section,numberless}[display]{}{}{0pt}{\normalfont\Huge\bfseries #1}
\fi

\usepackage{hyperref}    
\usepackage[all]{hypcap} 
\usepackage{cite} 
\usepackage{mciteplus}

\usepackage{multirow} 
\usepackage{booktabs} 

\usepackage{longtable} 
\usepackage{float}
\usepackage{amsmath}
\usepackage{tablefootnote}
\usepackage{comment}
\begin{document}

\newcommand{\xs}{0.95}
\newcommand{\xsstat}{0.04}
\newcommand{\xssyst}{0.08}
\newcommand{\xslumi}{0.02}
\newcommand{\tbarxs}{0.81}
\newcommand{\tbarxsstat}{0.03}
\newcommand{\tbarxssyst}{0.07}
\newcommand{\tbarxslumi}{0.02}
\newcommand{\ay}{0.08}
\newcommand{\aystat}{0.03}
\newcommand{\aysyst}{0.01}

\newcommand{\selmup}{0.23}
\newcommand{\selmum}{0.25}
\newcommand{\muonrecomup}{1.48}
\newcommand{\muonrecomum}{1.48}
\newcommand{\jetrecomup}{0.90}
\newcommand{\jetrecomum}{0.89}
\newcommand{\jesjermup}{4.96}
\newcommand{\jesjermum}{4.95}
\newcommand{\abcdmup}{2.03}
\newcommand{\abcdmum}{2.05}
\newcommand{\nlowbmup}{0.90}
\newcommand{\nlowbmum}{1.26}
\newcommand{\mistagmup}{5.82}
\newcommand{\mistagmum}{5.59}
\newcommand{\btaggingmup}{5.82}
\newcommand{\btaggingmum}{5.59}
\newcommand{\fitresultvalue}{0.743}
\newcommand{\fitresulterr}{0.013}
\newcommand{\TotalAsySigma}{2.64}
\def\asy {A_C^{t}}
\def\zpt         {\mbox{$p_{\mathrm{ T}}^Z$}\xspace}
\def\jeteta         {\mbox{$\eta_{\mathrm{jet}}$}\xspace}
\def\mupt  {\mbox{$p_{\mathrm{T,\, \mu}}$}\xspace}
\def\pttot {\mbox{$p_{\mathrm{T,\, total}}$}\xspace}
\def\jetpt {\mbox{$p_{\mathrm{T,\, jet}}$}\xspace}

\def\mueta         {\mbox{$\eta_{\mu}$}\xspace}
\renewcommand{\thefootnote}{\fnsymbol{footnote}}
\setcounter{footnote}{1}


\begin{titlepage}
\pagenumbering{roman}

\vspace*{-1.5cm}
\centerline{\large EUROPEAN ORGANIZATION FOR NUCLEAR RESEARCH (CERN)}
\vspace*{1.5cm}
\noindent
\begin{tabular*}{\linewidth}{lc@{\extracolsep{\fill}}r@{\extracolsep{0pt}}}
\ifthenelse{\boolean{pdflatex}}
{\vspace*{-1.5cm}\mbox{\!\!\!\includegraphics[width=.14\textwidth]{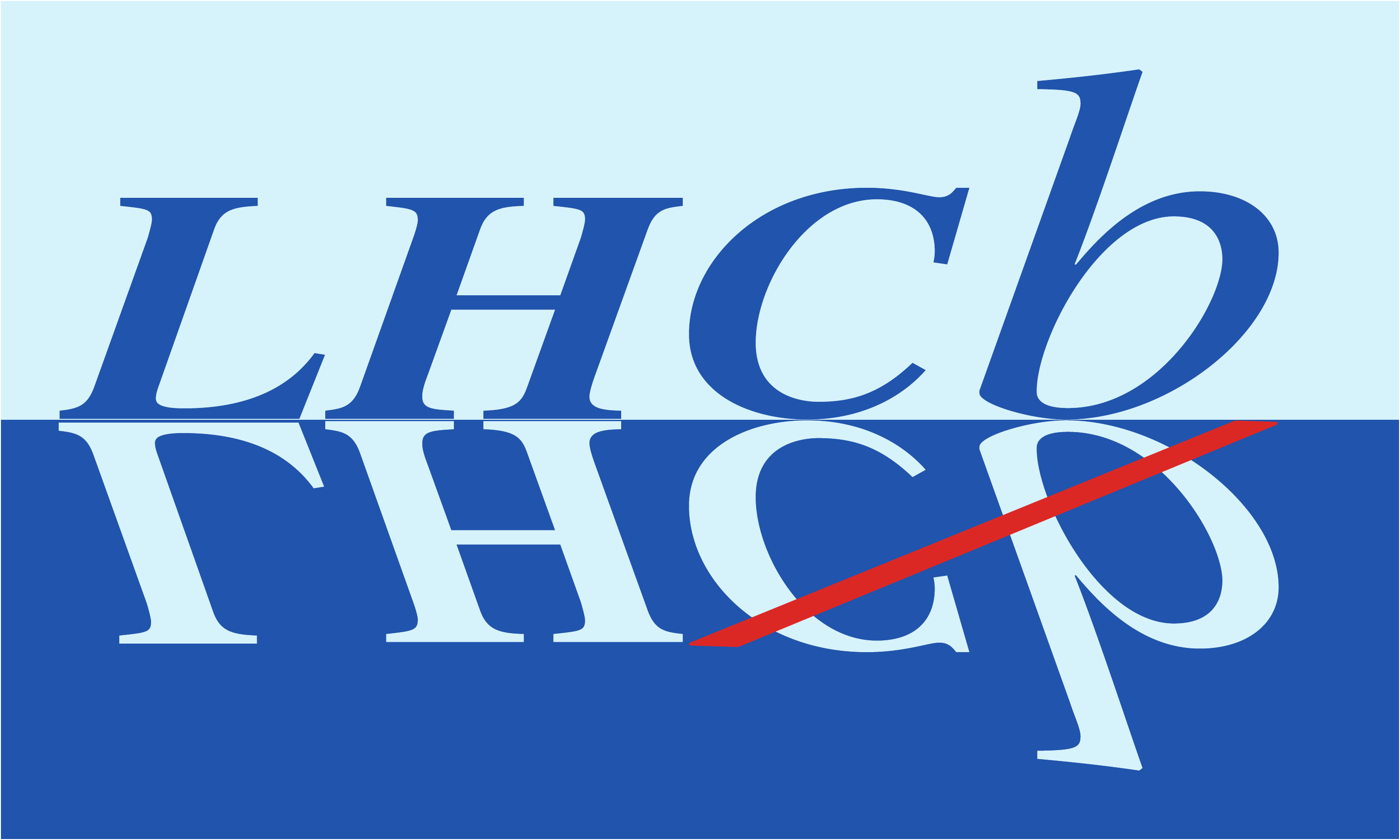}} & &}%
{\vspace*{-1.2cm}\mbox{\!\!\!\includegraphics[width=.12\textwidth]{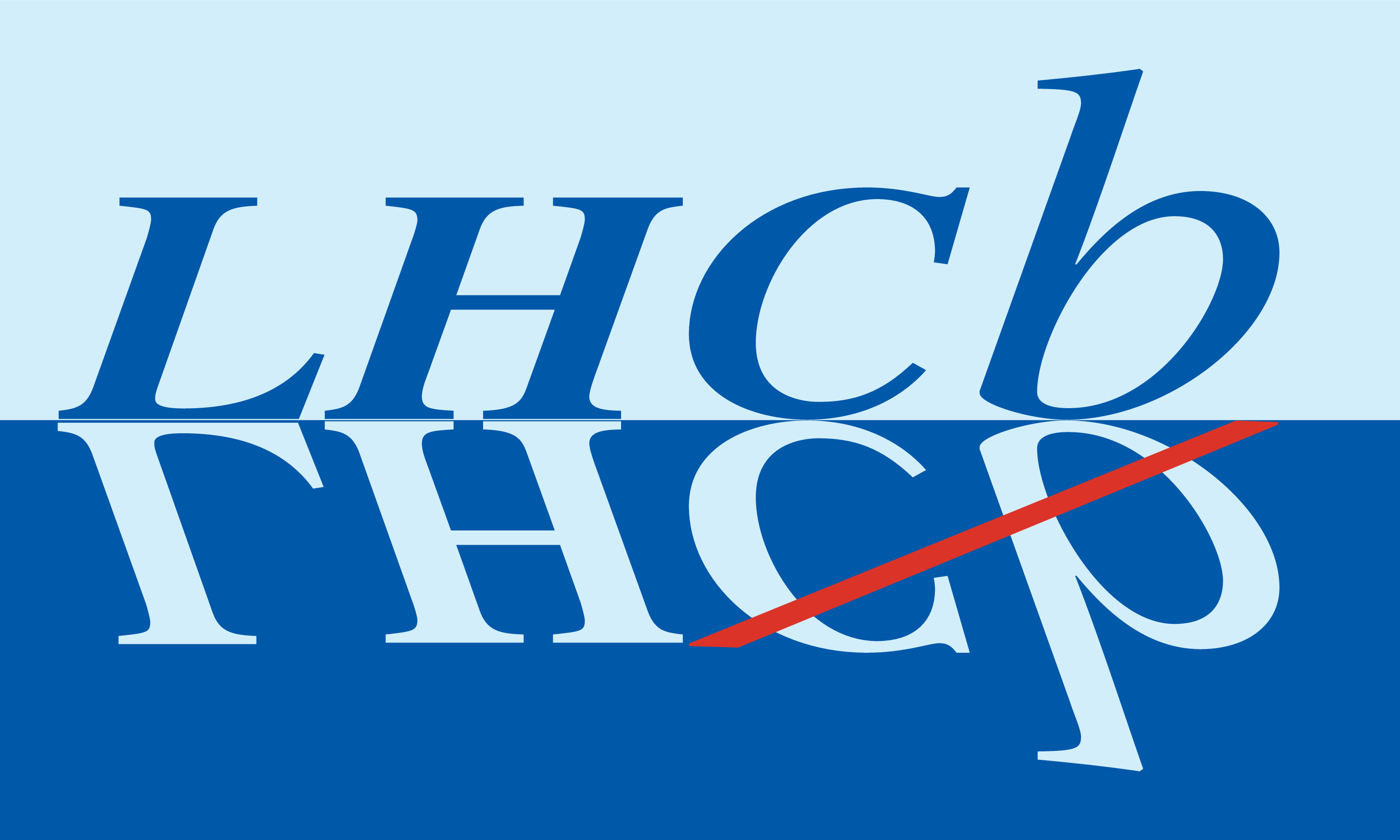}} & &}%
\\
 & & CERN-EP-2025-274 \\  
 & & LHCb-PAPER-2025-057 \\  
 & & June 1, 2026  \\ 
 & & \\
\end{tabular*}

\vspace*{1.50cm}

{\normalfont\bfseries\boldmath\huge
\begin{center}
\papertitle
\end{center}
}
\vspace*{1.5cm}
\begin{center}
\paperauthors\footnote{Authors are listed at the end of this paper.}
\end{center}

\vspace{\fill}

\begin{abstract}
  \noindent
The first measurements of the top- and antitop-quark differential production cross-sections and the top-quark charge asymmetry in the forward region are presented, using proton-proton collision data collected by the \lhcb experiment at a center-of-mass energy of 13\tev corresponding to an integrated luminosity of 5.4\invfb. The total production cross-sections of top and antitop quarks are also determined. 
Measurements are performed using the $\mu+\bjet$ final state within a fiducial region defined by \bjet $\jetpt>50\gev$ and pseudorapidity $2.2<\jeteta<4.0$, with the muon from the \W-boson decay required to have $\mupt>25\gev$ and $2.0<\mueta<4.5$. The muon and \bjet system must satisfy $\pt(\mu+\jet) > 20\gev$.
The measured integrated production cross-sections for the top and antitop quarks are
\begin{equation*}
\begin{split}
 \sigma_{\tquark} = \xs \pm \xsstat \pm \xssyst  \pm \xslumi \   \pb, \\
 \sigma_{\tquarkbar} = \tbarxs \pm \tbarxsstat  \pm \tbarxssyst  \pm \tbarxslumi \   \pb,
 \end{split}
\end{equation*}
 where the first uncertainty is statistical, the second systematic, and the third accounts for the luminosity uncertainty. The top-quark charge asymmetry is measured to be
\begin{equation*}
 \asy = \ay \pm \aystat  \pm \aysyst ,
\end{equation*}
 where the first uncertainty is statistical and the second is systematic. 
These results are consistent with next-to-leading order Standard Model predictions.
\end{abstract}
\vspace*{1.0cm}
\begin{center}
  Published in
  Phys.~Rev.~Lett. 136 (2026) 221801 
\end{center}

\vspace{\fill}

{\footnotesize 
\centerline{\copyright~\papercopyright. \href{\paperlicenceurl}{\paperlicence}.}}
\vspace*{2mm}

\end{titlepage}


\newpage
\setcounter{page}{2}
\mbox{~}
%
%
%
%

\renewcommand{\thefootnote}{\arabic{footnote}}
\setcounter{footnote}{0}


\cleardoublepage


\pagestyle{plain} 
\setcounter{page}{1}
\pagenumbering{arabic}


In the Standard Model (SM), the top quark (\tquark) is the most massive fundamental particle~\cite{PDG2024}. Owing to its large Yukawa coupling, it plays a central role in electroweak symmetry breaking and in interactions of the Higgs boson. Moreover, the top quark provides a sensitive probe of physics beyond the SM, including scenarios with modified Higgs couplings or new heavy states~\cite{Kagan:2011pnt,Gauld:2013aja,Gauld:2014pxa}.  At the Large Hadron Collider (\lhc), top quarks are mainly produced in pairs with their antiquarks (\ttbar) via the strong interaction, with a smaller contribution from single-top processes mediated by the electroweak interaction~\cite{PDG2024}. In the forward region, the SM predicts that about $80\%$ of $\tquark\to Wb$ decays originate from $\ttbar$ production, with the remaining $20\%$ dominated by $t$-channel single-top production~\cite{ttbar,ST_sch,Wt}.
The top-quark production cross-section is particularly sensitive to the gluon parton distribution function (PDF), especially in the high-Bjorken-$x$ region where current constraints remain weak~\cite{gluonpdf,Alekhin:2024bhs,Ablat:2023tiy}. 
With its unique forward acceptance, the \lhcb detector can probe kinematic regimes inaccessible to \atlas and \cms experiments, providing complementary measurements. In particular, forward top production has been suggested as a probe to constrain the gluon PDF at high $x$ at the level of $20\%$~\cite{Gauld:2014pxa}. 

Within the SM, \ttbar production is intrinsically charge symmetric at leading order, and the $gg\to \ttbar$ contribution is charge symmetric to all orders. However, next-to-leading-order (NLO) quantum chromodynamics (QCD) effects, driven by quark-antiquark annihilation, induce a small asymmetry. 
These effects arise from interference between initial- and final-state radiation, as well as between the Born and box diagrams~\cite{Bernreuther:2012sx}. 
Although single-top production has a small cross-section, its intrinsic asymmetry is enhanced in proton-proton ($pp$) collisions, where the larger \uquark-quark parton density relative to \dquark quarks leads to a net positive charge asymmetry; the single-top process itself yields a charge asymmetry at the level of about $40\%$~\cite{ST_sch,Wt}.
The charge-asymmetry observable is defined as
\begin{equation}
\label{eq:asy}
\asy = \left( \frac{\sigma_t - \sigma_{\bar{t}}}{\sigma_t + \sigma_{\bar{t}}} \right)_{\eta_{\mu}},
\end{equation}
where $\sigma_t$ and $\sigma_{\bar{t}}$ respectively represent the production cross-sections of \tquark and \tquarkbar quarks in bins of muon pseudorapidity, $\mueta$, 
measured in the \lhcb laboratory frame.
At the \lhc, where gluon fusion dominates, the charge asymmetry in \ttbar production is significantly diluted. Nevertheless, in the forward region accessible to \lhcb, this dilution is reduced, enhancing sensitivity to the charge asymmetry~\cite{Gauld:2014pxa}. Theoretical predictions suggest that \lhcb may provide the first significant observation of the top-quark charge asymmetry at the \lhc~\cite{Kagan:2011pnt,Gauld:2014pxa}. 

The \ttbar and single-top production cross-sections have been extensively measured by the \atlas and \cms collaborations at center-of-mass energies of $\sqs=5.02$, 7, 8, 13, and 13.6\tev~\cite{CMS:2012hcu,CMS:2012xhh,CMS:2013yjt,ATLAS:2014sxe,CMS:2014mgj,ATLAS:2014nxi,CMS:2015auz,ATLAS:2016qhd,CMS:2016yys,CMS:2016csa,ATLAS:2017jkf,ATLAS:2017wvi,ATLAS:2017rso,CMS:2018fks,CMS:2018lgn,CMS:2019snc,ATLAS:2020ccu,CMS:2021gwv,CMS:2021vhb,ATLAS:2022aof,ATLAS:2022jbj,ATLAS:2022opp,ATLAS:2023gsl,ATLAS:2023slx,CMS:2023qyl,CMS:2024ghc}. 
At \lhcb, previous studies have so far been limited to total cross-section measurements at 7, 8, and 13\tev~\cite{LHCb-PAPER-2015-022,LHCb-PAPER-2016-038,LHCb-PAPER-2017-050}.
Measurements of $\asy$ were first performed at the \tevatron by the \cdf and \dzero experiments~\cite{CDF:2011xdt,D0:2015alu,CDF:2012ctl}, where initial tensions with theoretical predictions were later reduced by improved calculations and updated measurements~\cite{Czakon:2014xsa,D0:2014cda,Kidonakis:2015ona,CDF:2017cvy}. At the \lhc, results from \atlas~\cite{ATLAS:2013buu,ATLAS:2015ysm,ATLAS:2015jgj,ATLAS:2015sex,ATLAS:2022wec,ATLAS:2022waa} and \cms~\cite{CMS:2012oht,CMS:2014rdf,CMS:2015fvy,CMS:2015pob,CMS:2016ypc} collaborations have achieved increasing precision.
While the most recent ATLAS measurement~\cite{ATLAS:2022waa} reports a 4.7$\sigma$ deviation from zero, the extracted asymmetry remains numerically small and consistent with the SM.

This Letter presents the first measurements of the differential production cross-sections of the \tquark and \tquarkbar quarks as a function of the muon pseudorapidity in the forward region, the corresponding $\asy$ observable, and the integrated production cross-sections.
The measurement is performed in the \decay{t}{W^{+}(\to\mu^{+}\nu_{\mu})\, b} decay channel,\footnote{Charge-conjugate states are implied throughout this Letter.} using $pp$ collision data collected with the \lhcb detector at $\sqs=13\tev$, corresponding to an integrated luminosity of 5.4\invfb.
The measurement is performed in the fiducial phase space defined by the presence of a muon with transverse momentum (defined with respect to the beam axis) $\mupt>25\gev$ in the pseudorapidity range $2.0<\mueta<4.5$.\footnote{Natural units with $\hbar = c = 1$ are used throughout.}
Jets are reconstructed using the anti-$k_\textrm{T}$ algorithm~\cite{antiKt} with radius parameter $R=0.5$, and are required to satisfy $\jetpt>50\gev$ and $2.2 < \jeteta < 4.0$. The inputs to the jet‑clustering algorithm are charged and neutral particles, which are obtained by combining information from all detector subsystems through a particle-flow approach~\cite{LHCb-PAPER-2013-058}.

The top charge asymmetry, $\asy$, is measured via Eq.~\ref{eq:asy}, while the differential cross-section for each charge of the final-state muon and each bin in $\eta_{\mu}$ is determined as
\begin{equation}
\label{eq:xsec}
\frac{d\sigma_{t}^{}}{d\eta_{\mu}} = 
\frac{\left( N - N_{\text{bkg}} \right) \cdot \mathcal{A} \cdot P}
{\Delta \eta_{\mu} \cdot \lum \cdot \varepsilon_{\text{sel}} \cdot \varepsilon_{\text{rec}} \cdot \varepsilon_{\text{btag}} \cdot \epsilon_{\text{true}} }, 
\end{equation}
where $N$ is the total number of selected candidates, $N_{\text{bkg}}$ is the estimated background contribution, and $\Delta \eta_{\mu}$ is the width of the bin in $\eta_{\mu}$. The integrated luminosity is denoted by \lum, and the  reconstruction, selection, and \bquark-tagging efficiencies are represented by $\varepsilon_{\text{rec}}$, $\varepsilon_{\text{sel}}$, and $\varepsilon_{\text{btag}}$, respectively. The acceptance factor, $\mathcal{A}$, accounts for migrations into and out of the fiducial region, including non-negligible contributions from the $\decay{t}{W^+(\to\tau^+\nu_{\tau})\, b}$ decay mode. The jet reconstruction purity, $P$, is defined as the fraction of reconstructed jets within the fiducial region that originate from truth-level jets also contained in the same region.  
The term $\epsilon_{\text{true}}$ denotes the truth-level jet reconstruction efficiency, defined as the fraction of truth-level jets within the fiducial region that are successfully reconstructed in the same region.  

The \lhcb detector~\cite{LHCb-DP-2008-001,LHCb-DP-2014-002} is a single-arm forward spectrometer covering $2<\eta <5$, designed for the study of particles containing \bquark or \cquark quarks. 
The detector used to collect data for this analysis includes a high-precision tracking system consisting of a silicon-strip vertex detector surrounding the $pp$ interaction region~\cite{LHCb-DP-2014-001}, a large-area silicon-strip detector located upstream of a dipole magnet with a bending power of about $4{\mathrm{\,T\,m}}$, and three stations of silicon-strip detectors and straw drift tubes~\cite{LHCb-DP-2017-001} placed downstream of the magnet.
The tracking system provides a measurement of the momentum, \ptot, of charged particles with a relative uncertainty that varies from 0.5\% at low momentum to 1.0\% at 200\gev. The minimum distance of a track to a primary $pp$ collision vertex (PV), the impact parameter (IP), is measured with a resolution of $(15+29/\pt)\mum$, where \pt is measured in\,\gev.
Different types of charged hadrons are distinguished using information from two ring-imaging Cherenkov detectors~\cite{LHCb-DP-2012-003}. 
Photons, electrons and hadrons are identified by a calorimeter system consisting of scintillating-pad and preshower detectors, an electromagnetic and a hadronic calorimeter.
Muons are identified by a system composed of alternating layers of iron and multiwire proportional chambers~\cite{LHCb-DP-2012-002}.
Approximately half of the data were collected with the magnet in each of the two polarity configurations. The online event selection is performed by a trigger system~\cite{LHCb-DP-2012-004,LHCb-DP-2019-001}. This analysis uses events selected by trigger algorithms that require at least one identified muon with high \pt.

Simulation samples are generated using \pythia~\cite{Sjostrand:2007gs,*Sjostrand:2006za} with a specific \lhcb configuration~\cite{LHCb-PROC-2010-056}. Final-state radiation is simulated with \photos~\cite{davidson2015photos}. The interaction of the generated particles with the detector, and its response, are modeled using the \geant toolkit~\cite{Allison:2006ve, *Agostinelli:2002hh} as described in Ref.~\cite{LHCb-PROC-2011-006}. 
Auxiliary signal samples without detector simulation are produced using the ‌\powheg-BOX framework~\cite{Nason:2004rx,Frixione:2007vw,Alioli:2010xd} at NLO, and interfaced with \pythia for parton showering and hadronization.
Alternative NLO predictions produced with the {\sc MadGraph}~\cite{Mg5} framework are employed for cross-checks.

Candidate \decay{t}{W^{+}(\to\mu^{+}\nu_{\mu})\, b} decays are formed from a well-separated muon-jet pair with $\Delta R(\mu, \jet) > 0.5$, where $\Delta R \equiv \sqrt{(\Delta \eta)^2 + (\Delta \phi)^2}$, and $\Delta \eta$ ($\Delta \phi$) is the difference in pseudorapidity (azimuthal angle) between the muon and the jet axis. 
To suppress semileptonic heavy-flavor backgrounds, the muon IP is required to be less than 0.04\mm~\cite{LHCb-PAPER-2016-011}. 
Hadrons misidentified as muons are suppressed by requiring the energy deposit in the electromagnetic $(E_{\rm{ECAL}})$ and hadronic $(E_{\rm{HCAL}})$ calorimeters to be small, $(E_{\rm ECAL}+E_{\rm HCAL})/p < 4\%$, where $p$ is the muon momentum.
The relative momentum uncertainty of the muon track, $\sigma_{p}/p$, is required to be less than 10\%.
Events containing a second, oppositely charged, high-\pt muon with $M(\mu^{+}\mu^{-})>40\gev$ are rejected to suppress $\Z/\gamma^{*}(\to\mu^+\mu^-)+\jet$ contamination.

The jet flavor is identified using a dedicated deep neural network (DNN) classifier, inspired by the DeepJet algorithm~\cite{deepjet}, and trained using simulated \bquark-, \cquark-, and light-flavor jet samples~\cite{LHCb-PAPER-2025-034}. The network uses more than 400 input features related to jet properties, including those associated with charged and neutral jet constituents, secondary vertex variables, and global features. 
The classifier's outputs are three probabilities, $P_b$, $P_c$, and $P_q$, which sum to unity and represent the likelihood of a jet originating from a $\bquark$-, $\cquark$-, or light-flavor parton, respectively.
To suppress contamination from \cquark-jets and light-flavor jets while maximizing the \bquark-jet signal significance, the optimal selection thresholds on the DNN outputs are determined by maximizing the expected signal significance, $S = N_\text{sim}/\sqrt{N_\text{data}}$, where $N_\text{sim}$ and $N_\text{data}$ denote the expected number of simulated top signal and data candidates passing the selection, respectively. 
The simulated yields are normalized using the theoretical predictions for the production cross-section and integrated luminosity.
The working point is chosen as $P_b > 0.65$ and $P_q < 0.05$.
Despite the tight $P_b$ requirement, residual contamination from misidentified \cquark-jets and light-flavor jets remains. To account for this, a template fit to the $P_b$ distribution in data is performed to extract the \bquark-jet fraction, using simulated templates for each jet flavor. The fit yields a \bquark-jet purity of approximately 74\%, which is applied as a bin-by-bin weight in $\eta_\mu$ to remove the contributions from \cquark-jets and light-flavor jets.

One of the dominant background contributions in this analysis arises from QCD multijet production, particularly from heavy-flavor decays or in-flight hadron decays producing high-\pt muons.  
In signal events such as \decay{t}{W^{+}(\to\mu^{+}\nu_{\mu})\, b}, the undetected neutrino carries a significant fraction of the \W-boson \pt, 
resulting in a strong momentum imbalance between the muon and the accompanying jet.  
In contrast, QCD events usually have the muon produced inside or near a jet, 
leading to a more balanced configuration where the muon and jet \pt are similar and the muon is less isolated. 
To suppress this background, candidates are required to satisfy two further kinematic criteria. 
First, the \pt of the $\mu + \bquark$-jet system, $\pttot = p_{\mathrm{T}}(j + j_{\mu})$, must be greater than 20\gev, where $j$ denotes the reconstructed jet and $j_{\mu}$ denotes the jet which contains the muon.  
Secondly, the muon isolation, defined as $\imu = \mupt /p_{\mathrm{T,}\, j_{\mu}}$, must exceed 0.9.
The residual QCD background is estimated using a data-driven \texttt{ABCD} method~\cite{ABCD_ATLAS} in the ($\pttot$, $\imu$) plane. In this analysis, the \texttt{ABCD} regions are defined as: region \texttt{A} with $\pttot < 15\gev$ and $\imu < 0.9$; region \texttt{B} with $\pttot < 15\gev$ and $\imu > 0.9$; region \texttt{C} with $\pttot > 20\gev$ and $\imu < 0.9$; and region \texttt{D} (the signal region) with $\pttot > 20\gev$ and $\imu > 0.9$, containing 2278 $\mu^+$+\bjet and 1799 $\mu^-$+\bjet events.
The estimated signal yield in region \texttt{D}, $N_{D}^{\text{sig}}$, is calculated using the generalized \texttt{ABCD} relation.

Electroweak backgrounds are also considered. The \Z + \bjet background contribution is estimated using $Z(\to \mumu)+ b$-jet data candidates together with an acceptance correction factor, taken from the ratio of yields in simulation. 
Here, the $Z(\to \mumu) + \bjet$ category corresponds to partially reconstructed $Z(\to \mumu) + \bjet$ decays where one muon falls outside the geometrical acceptance of the \lhcb detector.
This correction is applied in bins of $\eta_{\mu}$ and separately for each muon charge. After subtracting the QCD and $\Z+ \jet$ backgrounds, the yield of the remaining $\W+\bjet$ background contribution is estimated. The relative normalization between $\W+\bjet$ and $\W+\jet$ events is taken from the \powheg prediction and cross-checked with {\sc MadGraph}, with corrections for the \bquark-tagging efficiency applied to match the observed yields.

After background subtraction, the observed signal yields of 1458 (\tquark) and 1231 (\tquarkbar) are corrected for detector and selection effects. These include the muon reconstruction efficiency, the efficiency of the muon selection criteria, the jet reconstruction efficiency, the \bquark-tagging efficiency and mistag rates, possible bin-to-bin migrations due to the detector resolution, and finally the acceptance difference between the fiducial and signal definitions and a small contamination from $W$ bosons that decay via a $\tau$ lepton subsequently producing a muon. The muon reconstruction efficiency, encompassing track reconstruction, identification, and trigger requirements, is estimated with a tag-and-probe method using \Zmm data, following the procedures used in published \W- and \Z-boson cross-section measurements at \lhcb~\cite{LHCB-PAPER-2021-037,LHCB-PAPER-2014-033,LHCB-PAPER-2015-001,LHCB-PAPER-2015-049}. The muon reconstruction efficiency varies across the muon pseudorapidity region, ranging from 64\% to 77\%. The selection efficiencies are evaluated using simulated \ttbar and single-top samples, and vary across the muon pseudorapidity region, ranging from 68\% to 75\%.
The jet reconstruction efficiency is determined from simulated $\Z(\to\mu^+\mu^-)+\jet$ samples, and is found to be about 96\%. The \bquark-tagging efficiency, based on DNN discriminators, is evaluated using simulated \ttbar and single-top samples, and corrected with a data-driven tag-and-probe method in a dijet sample enriched in $b$-jets. The number of $b$-jets before applying the DNN requirement is determined from a fit to the $P_b$ template distribution. After the DNN selection, the mistag fraction, arising from $c$- and light-flavor jets, is subtracted to obtain the final $b$-jets yield. The mistag probability is evaluated as the fraction of simulated $c$-jets and light-flavor jets passing the same DNN selection under the same tag-and-probe conditions~\cite{LHCb-PAPER-2025-034}. 
After applying weights based on the ratio of data to simulation, found to vary between 0.94 and 1.04, the resulting \bquark-tagging efficiency is approximately 56\%. 

To account for detector-resolution effects, migration corrections for events with a reconstructed jet inside the fiducial region while the truth-level jet is outside, or vice versa, are evaluated with simulations.
The corrections from these effects are below 1\%. Bin-to-bin migrations in $\eta_\mu$ are found to be negligible, owing to the excellent resolution. 

Beyond detector-resolution effects, differences between the definitions of the fiducial region and of the signal region \texttt{D} (in the \texttt{ABCD} method) are taken into account. In the fiducial region, the \pt requirement is applied to the vector sum of the muon and the \bjet, while in the signal region it is applied to the muon-containing jet combined with the \bjet. This mismatch is corrected by using an acceptance factor defined as $\mathcal{A}_{j_{\mu}} =\frac{N\left( p_{\mathrm{T}}(\mu + j) > 20\,\gev \right)}{N\left( p_{\mathrm{T}}(j_{\mu} + j) > 20\,\gev \right)}$, which is evaluated from simulated top-quark events.
The data sample includes background contributions from muons originating in intermediate $\tau$ decays, in addition to those produced directly in \decay{W}{\mu\nu} decays. To ensure that the reported cross-sections correspond exclusively to prompt muons from \W decays, a correction factor, defined as the fraction of such muons relative to the total, is derived from simulation and applied to the yields measured in data. 
The effect of the acceptance factor is found to be below 2\%.

The sources of systematic uncertainties considered in the cross-section measurements are summarized in Table~\ref{tab:uncertainty}. 
The dominant contribution arises from \bquark-jet tagging, which comprises uncertainties from both the mistag fraction and the \bquark-tagging efficiency. These are affected by common factors such as the choice of the parton-shower model (comparing {\sc MadGraph} interfaced with \pythia and with {\sc Herwig}, with kinematic weighting applied), variations in the simulated flavor composition (achieved by altering the fractions of light-flavor and \cquark-jets in the fit), and the jet energy scale and resolution effects (evaluated through smearing and scaling). For the mistag fraction, an additional contribution arises from the statistical precision of the template fits. For the \bquark-tagging efficiency, the dominant statistical contribution comes from data-driven corrections, including the template fits and the weights based on the ratio of data to simulation.
The second-largest systematic uncertainty originates from the jet energy scale and resolution. The efficiency of the $\jetpt > 50\gev$ requirement is evaluated in simulation. A control sample of $Z+\jet$ events is used to assess the energy scale and resolution as done in Ref.~\cite{LHCb-PAPER-2020-018}, which are then propagated to the top-quark simulation to evaluate the impact.

\begin{table}
\begin{center}
\caption{Relative systematic uncertainties in the \tquark and \tquarkbar total cross-section measurements, in percent. The total uncertainties are obtained by summing in quadrature the statistical, total systematic, and luminosity components.}
\begin{tabular}{l|c|c}
\hline
 Source               & $\Delta\sigma_{\tquark}/\sigma_{\tquark}$ [\%]  & $\Delta\sigma_{\tquarkbar}/\sigma_{\tquarkbar}$ [\%] \\
\hline
\textit{b}-jet tagging & 6.5 & 6.2 \\
Jet energy scale and resolution & 5.0 & 5.0 \\ 
\texttt{ABCD} method  & 2.0 & 2.1 \\ 
$\Z+\bjet$ background & 0.2 & 0.2 \\ 
$\W+\bjet$ background & 0.9 & 1.3 \\ 
Selection efficiency & 0.2 & 0.2 \\ 
Muon reconstruction efficiency & 0.2 & 0.2 \\ 
Jet reconstruction efficiency & 0.3 & 0.3 \\ 
\hline
Total systematic (excl. lumi.) & 8.5 & 8.3 \\ 
Luminosity   & 2.0 & 2.0 \\ 
Statistical & 3.8 & 4.0 \\ 
Total  & 9.5 & 9.4 \\  \hline
\end{tabular}
\label{tab:uncertainty}
\end{center}
\end{table}

The uncertainty associated with the \texttt{ABCD} background estimation method receives contributions from several effects. The method assumes that the two discriminating variables are uncorrelated. To reduce possible correlations, the ratio $N_{C}/N_{A}$ in the region $\imu<0.9$ is calculated as a function of $\pt^{j_{\mu}}$, and used to weight both the \texttt{A} and \texttt{B} samples following the strategy of Ref.~\cite{LHCb-PAPER-2015-021}. The impact of this weighting is quantified by assessing the signal yield variations, with the difference assigned as the corresponding systematic uncertainty.
In addition, the impact of the uncertainties on the correction coefficients is evaluated by varying each coefficient independently by $\pm10\%$. The resulting variation in the extracted signal yield is taken as its systematic uncertainty.

The systematic uncertainty on the $\Z+\bjet$ background arises from the statistical uncertainty of the acceptance correction factor.
The uncertainty of the $\W+\bjet$ background originates from uncertainties in the theoretical prediction of the NLO cross-section ratio $\sigma(\W+\bjet)/\sigma(\W +\jet)$.
The effect of PDF uncertainties is determined using the technique of Monte Carlo PDF replicas, while the scale uncertainty is obtained from the envelope of variations where both the factorization and renormalization scales are independently varied up and down by a factor of two, as well as from the variation of the strong-coupling constant $\alpha_s$~\cite{theoryerr}.

Muon reconstruction uncertainties are derived from data-driven efficiency measurements using \Zmm events and applied to the simulated \tquark-quark samples. 
The jet reconstruction uncertainty is similarly evaluated using $\Z(\to\mu^{+}\mu^{-})+\jet$ control samples. The uncertainty from the selection efficiency accounts for the limited size of the simulated \tquark-quark samples. A 2\% uncertainty is assigned to the integrated luminosity, as determined in Ref.~\cite{lumi}.
The results obtained separately for the two magnet polarities are compatible within statistical uncertainties, and no additional systematic uncertainty is assigned.

The integrated production cross-sections for \tquark and \tquarkbar are measured using Eq.~\ref{eq:xsec} to be
\begin{equation*}
\begin{split}
 \sigma_{\tquark} = \xs \pm \xsstat  \stat \pm \xssyst \syst \pm \xslumi \ (\rm{lumi})  \pb, \\
 \sigma_{\tquarkbar} = \tbarxs \pm \tbarxsstat \stat \pm \tbarxssyst \syst \pm \tbarxslumi \ (\rm{lumi})  \pb,
 \end{split}
\end{equation*}
where the quoted uncertainties are statistical, systematic, and from the luminosity determination, respectively. The systematic uncertainties of the \tquark and \tquarkbar cross-sections are correlated at the level of 96\%. 
The inclusive charge asymmetry is measured as
\begin{equation*}
 \asy = \ay \pm \aystat \stat \pm \aysyst \syst.
\end{equation*}

\begin{figure}[!htbp]
    \centering
    \includegraphics[width=0.49\linewidth]{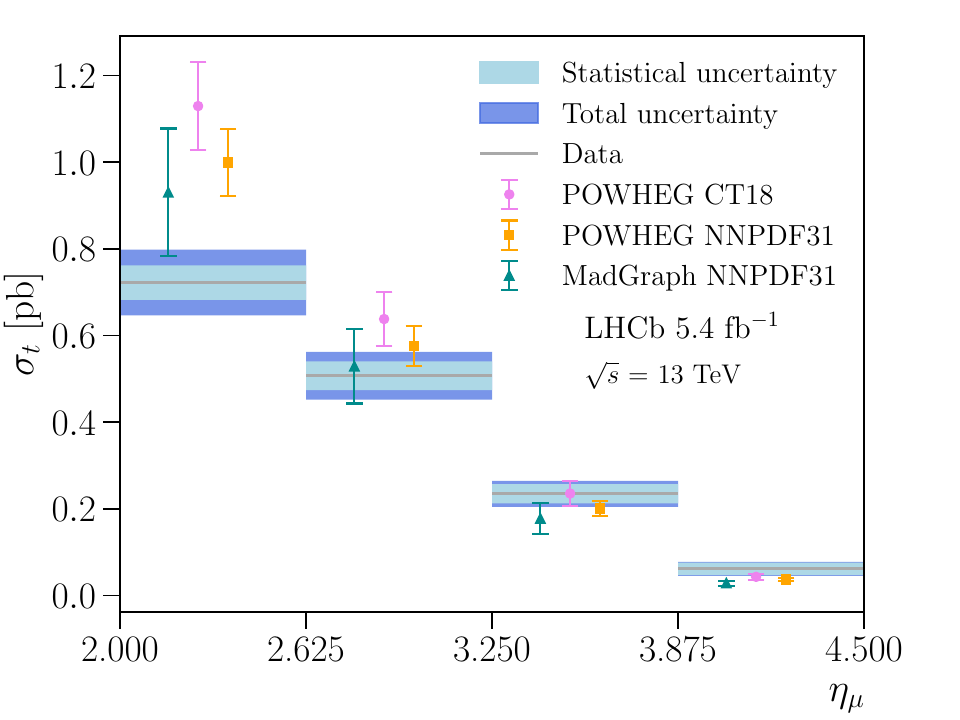}
    \includegraphics[width=0.49\linewidth]{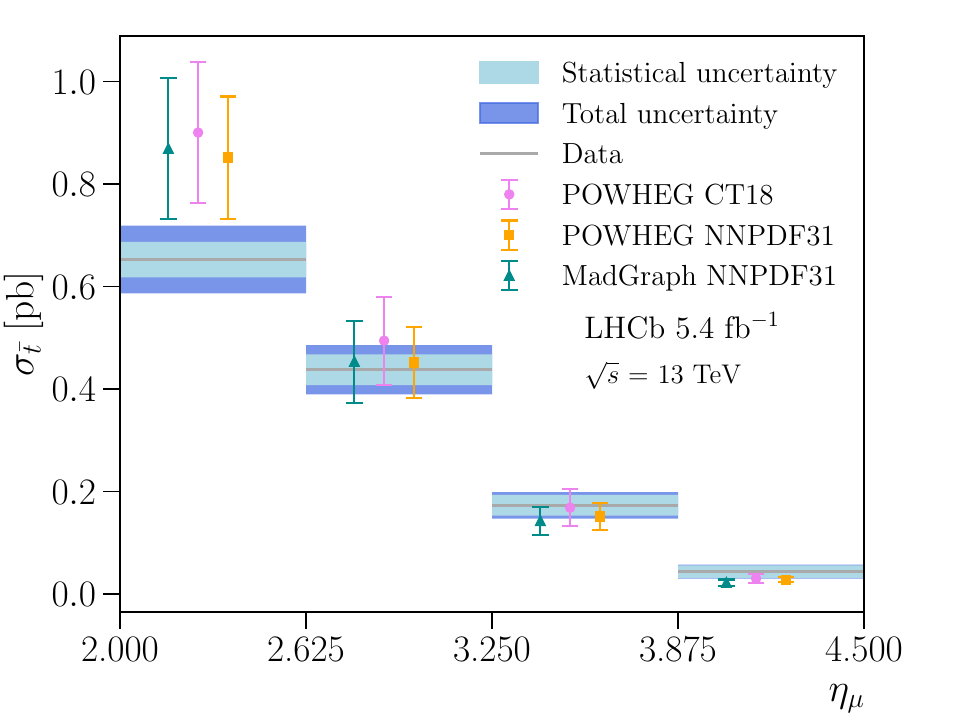}
    \caption{Differential cross-sections for the (left) \tquark and (right) \tquarkbar production as a function of the muon pseudorapidity compared with theoretical predictions at NLO.}
    \label{fig:xsec}
\end{figure}

\begin{figure}[!htbp]
    \centering
    \includegraphics[width=0.49\linewidth]{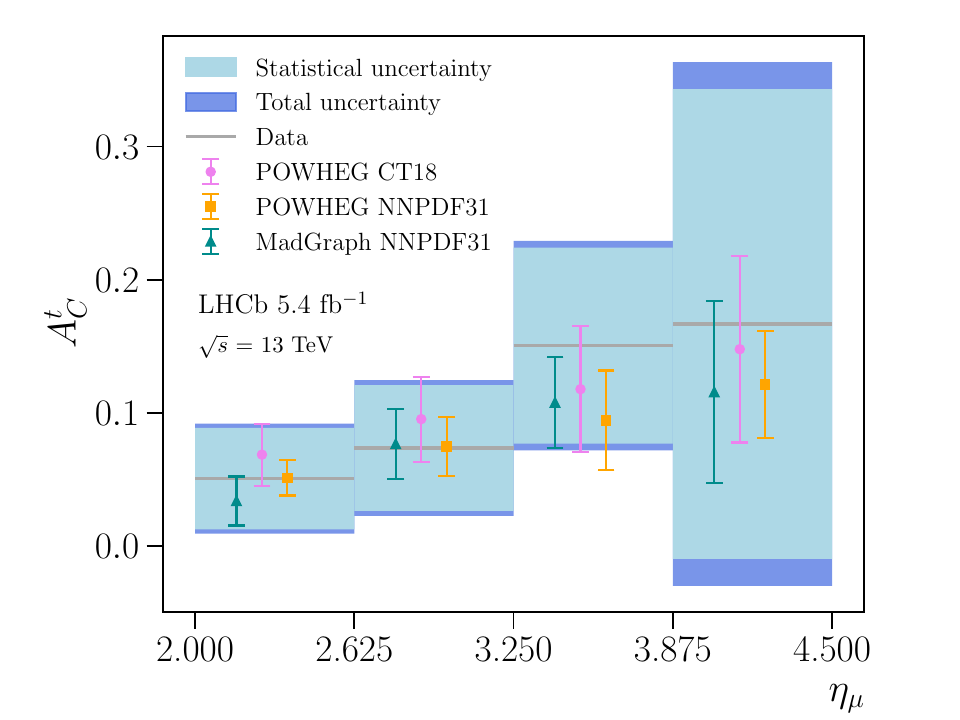}
    \caption{Measured \tquark-quark charge asymmetry as a function of the muon pseudorapidity compared with theoretical predictions at NLO.}
    \label{fig:asy}
\end{figure}

Differential cross-sections and charge asymmetries as a function of $\eta_\mu$ are shown in Figs.~\ref{fig:xsec} and \ref{fig:asy}, and summarized in Tables~\ref{tab:xsec_table} and \ref{tab:asymmetry_table}.
The measurements are compared with NLO predictions from \powheg-BOX~\cite{POWHEG} using CT18~\cite{Hou:2019efy} and NNPDF31~\cite{Ball2017} PDFs and {\sc MadGraph}~\cite{Mg5} using NNPDF31, and good agreement is observed.
Theoretical uncertainties contain the PDF, scale, and $\alpha_s$ uncertainties~\cite{bourilkov2006}. 

In summary, the first measurement of the top-quark charge asymmetry and the differential \tquark- and \tquarkbar-quark production cross-sections in the forward region is presented, using $pp$ collisions collected with the \lhcb detector between 2015 and 2018 at $\sqs=13\tev$, corresponding to an integrated luminosity of 5.4\invfb.
The results represent the most precise top-quark production cross-section measurements in the forward region to date.

For the charge asymmetry measurement, the $\mu+\bjet$ final state receives contributions from both $\ttbar$ and single-top production. In future analyses with larger data sets, the measured asymmetry should be decomposed through a fit incorporating the expected asymmetries from both processes. 

\emph{Data availability:} Data associated to the plots in this publication are made available on the CERN Document Server in Ref.~\cite{LHCb-PAPER-2025-057-cds}.

\begin{table}[!htbp]
    \centering
    \caption{Measured \tquark- and \tquarkbar-quark differential cross-sections as a function of $\eta_{\mu}$. Uncertainties are statistical, systematic, and from the luminosity, respectively.}
\begin{tabular}{ccc}
    \toprule
    $\eta_{\mu}$ interval & $\sigma_{\tquark}$ (pb) & $\sigma_{\tquarkbar}$ (pb) \\
    \midrule
    2.000--2.625 & $0.722 \pm 0.040 \pm 0.063 \pm 0.014$ & $0.653 \pm 0.035 \pm 0.055 \pm 0.013$ \\
    2.625--3.250 & $0.507 \pm 0.033 \pm 0.043 \pm 0.010$ & $0.437 \pm 0.030 \pm 0.037 \pm 0.009$ \\
    3.250--3.875 & $0.234 \pm 0.022 \pm 0.019 \pm 0.005$ & $0.173 \pm 0.020 \pm 0.016 \pm 0.003$ \\
    3.875--4.500 & $0.061 \pm 0.014 \pm 0.008 \pm 0.001$ & $0.044 \pm 0.012 \pm 0.007 \pm 0.001$ \\

    \bottomrule
\end{tabular}
    \label{tab:xsec_table}
\end{table}

\begin{table}[!htbp]
    \centering
    \caption{Measured \tquark-quark charge asymmetry as a function of $\eta_{\mu}$. Uncertainties are statistical and systematic, respectively. The significance quantifies the deviation from zero.}
\begin{tabular}{cccc}
    \toprule
    $\eta_{\mu}$ range & $\asy$ & significance  \\
    \midrule

    2.000--2.625 & $0.05 \pm 0.04 \pm 0.02$ & $1.22$ \\
    2.625--3.250 & $0.07 \pm 0.05 \pm 0.02$ & $1.44$ \\
    3.250--3.875 & $0.15 \pm 0.07 \pm 0.03$ & $1.91$ \\
    3.875--4.500 & $0.17 \pm 0.18 \pm 0.09$ & $0.85$ \\ \hline

    2.0--4.5\;\;(\text{combined}) 
      & $\ay \pm \aystat \pm \aysyst$ 
      & \TotalAsySigma \\

    \bottomrule
\end{tabular}
    \label{tab:asymmetry_table}
\end{table}

\clearpage
\section*{Acknowledgements}
%
%
\noindent We express our gratitude to our colleagues in the CERN
accelerator departments for the excellent performance of the LHC. We
thank the technical and administrative staff at the LHCb
institutes.
We thank James Vincent Mead and Stephen Farry
for their pioneering work on top-quark production, which has substantially informed
and supported the development of this measurement~\cite{james_paper}.
We acknowledge support from CERN and from the national agencies:
ARC (Australia);
CAPES, CNPq, FAPERJ and FINEP (Brazil); 
MOST and NSFC (China); 
CNRS/IN2P3 (France); 
BMFTR, DFG and MPG (Germany);
INFN (Italy); 
NWO (Netherlands); 
MNiSW and NCN (Poland); 
MCID/IFA (Romania); 
MICIU and AEI (Spain);
SNSF and SER (Switzerland); 
NASU (Ukraine); 
STFC (United Kingdom); 
DOE NP and NSF (USA).
We acknowledge the computing resources that are provided by ARDC (Australia), 
CBPF (Brazil),
CERN, 
IHEP and LZU (China),
IN2P3 (France), 
KIT and DESY (Germany), 
INFN (Italy), 
SURF (Netherlands),
Polish WLCG (Poland),
IFIN-HH (Romania), 
PIC (Spain), CSCS (Switzerland), 
and GridPP (United Kingdom).
We are indebted to the communities behind the multiple open-source
software packages on which we depend.
Individual groups or members have received support from
Key Research Program of Frontier Sciences of CAS, CAS PIFI, CAS CCEPP, 
Minciencias (Colombia);
EPLANET, Marie Sk\l{}odowska-Curie Actions, ERC and NextGenerationEU (European Union);
A*MIDEX, ANR, IPhU and Labex P2IO, and R\'{e}gion Auvergne-Rh\^{o}ne-Alpes (France);
Alexander-von-Humboldt Foundation (Germany);
ICSC (Italy); 
Severo Ochoa and Mar\'ia de Maeztu Units of Excellence, GVA, XuntaGal, GENCAT, InTalent-Inditex and Prog.~Atracci\'on Talento CM (Spain);
SRC (Sweden);
the Leverhulme Trust, the Royal Society and UKRI (United Kingdom).

\clearpage
\clearpage
\section*{End Matter}
\label{sec:Appendix}

For the differential cross-section measurement, the correlation matrix is constructed by evaluating the shift in each bin associated with each systematic variation. Uncertainties related to muon identification, tracking, trigger efficiencies, jet reconstruction efficiency, event selection efficiency, and the statistical components of the $b$-jet tagging efficiency are treated using the Root-Mean-Square (RMS) method, which quantifies their statistical spread by computing the root-mean-square of the resulting cross-section shifts across all variations. For uncertainties arising from the \texttt{ABCD} method, background estimation method, the jet energy scale and resolution and the components of the $b$-jet tagging systematics whose variations are intrinsic to the methodology itself, the envelope method is employed. In this approach, the largest absolute deviation among all variations is taken as the systematic uncertainty, and the correlations across bins are assumed to be 100\%. The correlation matrix for the differential cross-section measurements is presented in Table~\ref{tab:corr_mum_mup}. Owing to the limited statistics in the last bin, the systematic uncertainties have large fluctuations. This statistical limitation leads to the appearance of the small correlation in the correlation matrix.

\begin{table}[!htbp]
\caption{Correlation matrix for the systematic uncertainties in the differential cross-section measurements of \tquark\ and \tquarkbar quarks as a function of $\eta_{\mu}$. Bins 1–4 correspond to \tquarkbar\ ($\eta_{\mu^-}$) and bins 5–8 to \tquark\ ($\eta_{\mu^+}$). The luminosity uncertainty, fully correlated across all bins, is excluded.}
\centering
\begin{tabular}{c|cccccccc} 
 & \multicolumn{4}{c|}{$\sigma_{\bar{t}}$ ($\eta_{\mu^-}$)} 
 & \multicolumn{4}{c}{$\sigma_{t}$ ($\eta_{\mu^+}$)} \\     
\cline{2-9} 
Bin index & 1 & 2 & 3 & 4 & 5 & 6 & 7 & 8 \\
1  & $\phantom{-}1.00$ &  &  &  &  &  &  &  \\
2  & $\phantom{-}0.88$ & $\phantom{-}1.00$ &  &  &  &  &  &  \\
3  & $\phantom{-}0.83$ & $\phantom{-}0.68$ & $\phantom{-}1.00$ &  &  &  &  &  \\
4  & $\phantom{-}0.27$ & $\phantom{-}0.48$ & $\phantom{-}0.59$ & $\phantom{-}1.00$ &  &  &  &  \\
5  & $\phantom{-}0.93$ & $\phantom{-}0.98$ & $\phantom{-}0.73$ & $\phantom{-}0.42$ & $\phantom{-}1.00$ &  &  &  \\
6  & $\phantom{-}0.97$ & $\phantom{-}0.89$ & $\phantom{-}0.90$ & $\phantom{-}0.44$ & $\phantom{-}0.93$ & $\phantom{-}1.00$ &  &  \\
7  & $\phantom{-}0.84$ & $\phantom{-}0.96$ & $\phantom{-}0.79$ & $\phantom{-}0.67$ & $\phantom{-}0.93$ & $\phantom{-}0.91$ & $\phantom{-}1.00$ &  \\
8  & $\phantom{-}0.77$ & $\phantom{-}0.90$ & $\phantom{-}0.42$ & $\phantom{-}0.16$ & $\phantom{-}0.88$ & $\phantom{-}0.75$ & $\phantom{-}0.79$ & $\phantom{-}1.00$ \\
\hline
\end{tabular}
\label{tab:corr_mum_mup}
\end{table}

\begin{mcitethebibliography}{10}
\mciteSetBstSublistMode{n}
\mciteSetBstMaxWidthForm{subitem}{\alph{mcitesubitemcount})}
\mciteSetBstSublistLabelBeginEnd{\mcitemaxwidthsubitemform\space}
{\relax}{\relax}

\bibitem{PDG2024}
Particle Data Group, S.~Navas {\em et~al.}, \ifthenelse{\boolean{articletitles}}{\emph{{\href{http://pdg.lbl.gov/}{Review of particle physics}}}, }{}\href{https://doi.org/10.1103/PhysRevD.110.030001}{Phys.\ Rev.\  \textbf{D110} (2024) 030001}\relax
\mciteBstWouldAddEndPuncttrue
\mciteSetBstMidEndSepPunct{\mcitedefaultmidpunct}
{\mcitedefaultendpunct}{\mcitedefaultseppunct}\relax
\EndOfBibitem
\bibitem{Kagan:2011pnt}
A.~L. Kagan, J.~F. Kamenik, G.~Perez, and S.~Stone, \ifthenelse{\boolean{articletitles}}{\emph{{Probing new top physics at the LHCb experiment}}, }{}\href{https://doi.org/10.1103/PhysRevLett.107.082003}{Phys.\ Rev.\ Lett.\  \textbf{107} (2011) 082003}, \href{http://arxiv.org/abs/1103.3747}{{\normalfont\ttfamily arXiv:1103.3747}}\relax
\mciteBstWouldAddEndPuncttrue
\mciteSetBstMidEndSepPunct{\mcitedefaultmidpunct}
{\mcitedefaultendpunct}{\mcitedefaultseppunct}\relax
\EndOfBibitem
\bibitem{Gauld:2013aja}
R.~Gauld, \ifthenelse{\boolean{articletitles}}{\emph{{Feasibility of top quark measurements at LHCb and constraints on the large-$x$ gluon PDF}}, }{}\href{https://doi.org/10.1007/JHEP02(2014)126}{JHEP \textbf{02} (2014) 126}, \href{http://arxiv.org/abs/1311.1810}{{\normalfont\ttfamily arXiv:1311.1810}}\relax
\mciteBstWouldAddEndPuncttrue
\mciteSetBstMidEndSepPunct{\mcitedefaultmidpunct}
{\mcitedefaultendpunct}{\mcitedefaultseppunct}\relax
\EndOfBibitem
\bibitem{Gauld:2014pxa}
R.~Gauld, \ifthenelse{\boolean{articletitles}}{\emph{{Leptonic top quark asymmetry predictions at LHCb}}, }{}\href{https://doi.org/10.1103/PhysRevD.91.054029}{Phys.\ Rev.\  \textbf{D91} (2015) 054029}, \href{http://arxiv.org/abs/1409.8631}{{\normalfont\ttfamily arXiv:1409.8631}}\relax
\mciteBstWouldAddEndPuncttrue
\mciteSetBstMidEndSepPunct{\mcitedefaultmidpunct}
{\mcitedefaultendpunct}{\mcitedefaultseppunct}\relax
\EndOfBibitem
\bibitem{ttbar}
J.~M. Campbell, R.~K. Ellis, P.~Nason, and E.~Re, \ifthenelse{\boolean{articletitles}}{\emph{{Top-Pair Production and Decay at NLO Matched with Parton Showers}}, }{}\href{https://doi.org/10.1007/JHEP04(2015)114}{JHEP \textbf{04} (2015) 114}, \href{http://arxiv.org/abs/1412.1828}{{\normalfont\ttfamily arXiv:1412.1828}}\relax
\mciteBstWouldAddEndPuncttrue
\mciteSetBstMidEndSepPunct{\mcitedefaultmidpunct}
{\mcitedefaultendpunct}{\mcitedefaultseppunct}\relax
\EndOfBibitem
\bibitem{ST_sch}
S.~Alioli, P.~Nason, C.~Oleari, and E.~Re, \ifthenelse{\boolean{articletitles}}{\emph{{NLO single-top production matched with shower in POWHEG: s- and t-channel contributions}}, }{}\href{https://doi.org/10.1088/1126-6708/2009/09/111}{JHEP \textbf{09} (2009) 111}, \href{http://arxiv.org/abs/0907.4076}{{\normalfont\ttfamily arXiv:0907.4076}}\relax
\mciteBstWouldAddEndPuncttrue
\mciteSetBstMidEndSepPunct{\mcitedefaultmidpunct}
{\mcitedefaultendpunct}{\mcitedefaultseppunct}\relax
\EndOfBibitem
\bibitem{Wt}
E.~Re, \ifthenelse{\boolean{articletitles}}{\emph{{Single-top Wt-channel production matched with parton showers using the POWHEG method}}, }{}\href{https://doi.org/10.1140/epjc/s10052-011-1547-z}{Eur.\ Phys.\ J.\  \textbf{C71} (2011) 1547}, \href{http://arxiv.org/abs/1009.2450}{{\normalfont\ttfamily arXiv:1009.2450}}\relax
\mciteBstWouldAddEndPuncttrue
\mciteSetBstMidEndSepPunct{\mcitedefaultmidpunct}
{\mcitedefaultendpunct}{\mcitedefaultseppunct}\relax
\EndOfBibitem
\bibitem{gluonpdf}
M.~Czakon, M.~L. Mangano, A.~Mitov, and J.~Rojo, \ifthenelse{\boolean{articletitles}}{\emph{{Constraints on the gluon PDF from top quark pair production at hadron colliders}}, }{}\href{https://doi.org/10.1007/JHEP07(2013)167}{JHEP \textbf{07} (2013) 167}, \href{http://arxiv.org/abs/1303.7215}{{\normalfont\ttfamily arXiv:1303.7215}}\relax
\mciteBstWouldAddEndPuncttrue
\mciteSetBstMidEndSepPunct{\mcitedefaultmidpunct}
{\mcitedefaultendpunct}{\mcitedefaultseppunct}\relax
\EndOfBibitem
\bibitem{Alekhin:2024bhs}
S.~Alekhin, M.~V. Garzelli, S.-O. Moch, and O.~Zenaiev, \ifthenelse{\boolean{articletitles}}{\emph{{NNLO PDFs driven by top-quark data}}, }{}\href{https://doi.org/10.1140/epjc/s10052-025-13832-8}{Eur.\ Phys.\ J.\  \textbf{C85} (2025) 162}, \href{http://arxiv.org/abs/2407.00545}{{\normalfont\ttfamily arXiv:2407.00545}}\relax
\mciteBstWouldAddEndPuncttrue
\mciteSetBstMidEndSepPunct{\mcitedefaultmidpunct}
{\mcitedefaultendpunct}{\mcitedefaultseppunct}\relax
\EndOfBibitem
\bibitem{Ablat:2023tiy}
A.~Ablat {\em et~al.}, \ifthenelse{\boolean{articletitles}}{\emph{{Exploring the impact of high-precision top-quark pair production data on the structure of the proton at the LHC}}, }{}\href{https://doi.org/10.1103/PhysRevD.109.054027}{Phys.\ Rev.\  \textbf{D109} (2024) 054027}, \href{http://arxiv.org/abs/2307.11153}{{\normalfont\ttfamily arXiv:2307.11153}}\relax
\mciteBstWouldAddEndPuncttrue
\mciteSetBstMidEndSepPunct{\mcitedefaultmidpunct}
{\mcitedefaultendpunct}{\mcitedefaultseppunct}\relax
\EndOfBibitem
\bibitem{Bernreuther:2012sx}
W.~Bernreuther and Z.-G. Si, \ifthenelse{\boolean{articletitles}}{\emph{{Top quark and leptonic charge asymmetries for the Tevatron and LHC}}, }{}\href{https://doi.org/10.1103/PhysRevD.86.034026}{Phys.\ Rev.\  \textbf{D86} (2012) 034026}, \href{http://arxiv.org/abs/1205.6580}{{\normalfont\ttfamily arXiv:1205.6580}}\relax
\mciteBstWouldAddEndPuncttrue
\mciteSetBstMidEndSepPunct{\mcitedefaultmidpunct}
{\mcitedefaultendpunct}{\mcitedefaultseppunct}\relax
\EndOfBibitem
\bibitem{CMS:2012hcu}
CMS collaboration, S.~Chatrchyan {\em et~al.}, \ifthenelse{\boolean{articletitles}}{\emph{{Measurement of the $t\bar{t}$ production cross section in $pp$ collisions at $\sqrt{s}=7$ TeV with lepton + jets final states}}, }{}\href{https://doi.org/10.1016/j.physletb.2013.02.021}{Phys.\ Lett.\  \textbf{B720} (2013) 83}, \href{http://arxiv.org/abs/1212.6682}{{\normalfont\ttfamily arXiv:1212.6682}}\relax
\mciteBstWouldAddEndPuncttrue
\mciteSetBstMidEndSepPunct{\mcitedefaultmidpunct}
{\mcitedefaultendpunct}{\mcitedefaultseppunct}\relax
\EndOfBibitem
\bibitem{CMS:2012xhh}
CMS collaboration, S.~Chatrchyan {\em et~al.}, \ifthenelse{\boolean{articletitles}}{\emph{{Measurement of the single-top-quark $t$-channel cross section in $pp$ collisions at $\sqrt{s}=7$ TeV}}, }{}\href{https://doi.org/10.1007/JHEP12(2012)035}{JHEP \textbf{12} (2012) 035}, \href{http://arxiv.org/abs/1209.4533}{{\normalfont\ttfamily arXiv:1209.4533}}\relax
\mciteBstWouldAddEndPuncttrue
\mciteSetBstMidEndSepPunct{\mcitedefaultmidpunct}
{\mcitedefaultendpunct}{\mcitedefaultseppunct}\relax
\EndOfBibitem
\bibitem{CMS:2013yjt}
CMS collaboration, S.~Chatrchyan {\em et~al.}, \ifthenelse{\boolean{articletitles}}{\emph{{Measurement of the $t\bar{t}$ production cross section in the all-jet final state in pp collisions at $\sqrt{s}$ = 7 TeV}}, }{}\href{https://doi.org/10.1007/JHEP05(2013)065}{JHEP \textbf{05} (2013) 065}, \href{http://arxiv.org/abs/1302.0508}{{\normalfont\ttfamily arXiv:1302.0508}}\relax
\mciteBstWouldAddEndPuncttrue
\mciteSetBstMidEndSepPunct{\mcitedefaultmidpunct}
{\mcitedefaultendpunct}{\mcitedefaultseppunct}\relax
\EndOfBibitem
\bibitem{ATLAS:2014sxe}
ATLAS collaboration, G.~Aad {\em et~al.}, \ifthenelse{\boolean{articletitles}}{\emph{{Comprehensive measurements of $t$-channel single top-quark production cross sections at $\sqrt{s} = 7$ TeV with the ATLAS detector}}, }{}\href{https://doi.org/10.1103/PhysRevD.90.112006}{Phys.\ Rev.\  \textbf{D90} (2014) 112006}, \href{http://arxiv.org/abs/1406.7844}{{\normalfont\ttfamily arXiv:1406.7844}}\relax
\mciteBstWouldAddEndPuncttrue
\mciteSetBstMidEndSepPunct{\mcitedefaultmidpunct}
{\mcitedefaultendpunct}{\mcitedefaultseppunct}\relax
\EndOfBibitem
\bibitem{CMS:2014mgj}
CMS collaboration, V.~Khachatryan {\em et~al.}, \ifthenelse{\boolean{articletitles}}{\emph{{Measurement of the t-channel single-top-quark production cross section and of the $\mid V_{tb} \mid$ CKM matrix element in pp collisions at $\sqrt{s}$= 8 TeV}}, }{}\href{https://doi.org/10.1007/JHEP06(2014)090}{JHEP \textbf{06} (2014) 090}, \href{http://arxiv.org/abs/1403.7366}{{\normalfont\ttfamily arXiv:1403.7366}}\relax
\mciteBstWouldAddEndPuncttrue
\mciteSetBstMidEndSepPunct{\mcitedefaultmidpunct}
{\mcitedefaultendpunct}{\mcitedefaultseppunct}\relax
\EndOfBibitem
\bibitem{ATLAS:2014nxi}
ATLAS collaboration, G.~Aad {\em et~al.}, \ifthenelse{\boolean{articletitles}}{\emph{{Measurement of the $t\bar{t}$ production cross-section using $e\mu $ events with b-tagged jets in pp collisions at $\sqrt{s}$ = 7 and 8 $\,\mathrm{TeV}$ with the ATLAS detector}}, }{}\href{https://doi.org/10.1140/epjc/s10052-016-4501-2}{Eur.\ Phys.\ J.\  \textbf{C74} (2014) 3109}, Addendum \href{https://doi.org/10.1140/epjc/s10052-016-4501-2}{ibid.\   \textbf{76} (2016) 642}, \href{http://arxiv.org/abs/1406.5375}{{\normalfont\ttfamily arXiv:1406.5375}}\relax
\mciteBstWouldAddEndPuncttrue
\mciteSetBstMidEndSepPunct{\mcitedefaultmidpunct}
{\mcitedefaultendpunct}{\mcitedefaultseppunct}\relax
\EndOfBibitem
\bibitem{CMS:2015auz}
CMS collaboration, V.~Khachatryan {\em et~al.}, \ifthenelse{\boolean{articletitles}}{\emph{{Measurement of the $t\bar{t}$ production cross section in the all-jets final state in pp collisions at $\sqrt{s}=8$ $\,\text {TeV}$}}, }{}\href{https://doi.org/10.1140/epjc/s10052-016-3956-5}{Eur.\ Phys.\ J.\  \textbf{C76} (2016) 128}, \href{http://arxiv.org/abs/1509.06076}{{\normalfont\ttfamily arXiv:1509.06076}}\relax
\mciteBstWouldAddEndPuncttrue
\mciteSetBstMidEndSepPunct{\mcitedefaultmidpunct}
{\mcitedefaultendpunct}{\mcitedefaultseppunct}\relax
\EndOfBibitem
\bibitem{ATLAS:2016qhd}
ATLAS collaboration, M.~Aaboud {\em et~al.}, \ifthenelse{\boolean{articletitles}}{\emph{{Measurement of the inclusive cross-sections of single top-quark and top-antiquark $t$-channel production in $pp$ collisions at $\sqrt{s}$ = 13 TeV with the ATLAS detector}}, }{}\href{https://doi.org/10.1007/JHEP04(2017)086}{JHEP \textbf{04} (2017) 086}, \href{http://arxiv.org/abs/1609.03920}{{\normalfont\ttfamily arXiv:1609.03920}}\relax
\mciteBstWouldAddEndPuncttrue
\mciteSetBstMidEndSepPunct{\mcitedefaultmidpunct}
{\mcitedefaultendpunct}{\mcitedefaultseppunct}\relax
\EndOfBibitem
\bibitem{CMS:2016yys}
CMS collaboration, V.~Khachatryan {\em et~al.}, \ifthenelse{\boolean{articletitles}}{\emph{{Measurement of the $\ttbar$ production cross section in the $e\mu$ channel in proton-proton collisions at $\sqrt{s} =$ 7 and 8 TeV}}, }{}\href{https://doi.org/10.1007/JHEP08(2016)029}{JHEP \textbf{08} (2016) 029}, \href{http://arxiv.org/abs/1603.02303}{{\normalfont\ttfamily arXiv:1603.02303}}\relax
\mciteBstWouldAddEndPuncttrue
\mciteSetBstMidEndSepPunct{\mcitedefaultmidpunct}
{\mcitedefaultendpunct}{\mcitedefaultseppunct}\relax
\EndOfBibitem
\bibitem{CMS:2016csa}
CMS collaboration, V.~Khachatryan {\em et~al.}, \ifthenelse{\boolean{articletitles}}{\emph{{Measurements of the $t\bar{t}$ production cross section in lepton+jets final states in pp collisions at 8 $\,\text {TeV}$ and ratio of 8 to 7 $\,\text {TeV}$ cross sections}}, }{}\href{https://doi.org/10.1140/epjc/s10052-016-4504-z}{Eur.\ Phys.\ J.\  \textbf{C77} (2017) 15}, \href{http://arxiv.org/abs/1602.09024}{{\normalfont\ttfamily arXiv:1602.09024}}\relax
\mciteBstWouldAddEndPuncttrue
\mciteSetBstMidEndSepPunct{\mcitedefaultmidpunct}
{\mcitedefaultendpunct}{\mcitedefaultseppunct}\relax
\EndOfBibitem
\bibitem{ATLAS:2017jkf}
ATLAS collaboration, M.~Aaboud {\em et~al.}, \ifthenelse{\boolean{articletitles}}{\emph{{Measurement of the $t\bar{t}$ production cross section in the $\tau$ + jets final state in $pp$ collisions at $\sqrt{s}=8$ TeV using the ATLAS detector}}, }{}\href{https://doi.org/10.1103/PhysRevD.95.072003}{Phys.\ Rev.\  \textbf{D95} (2017) 072003}, \href{http://arxiv.org/abs/1702.08839}{{\normalfont\ttfamily arXiv:1702.08839}}\relax
\mciteBstWouldAddEndPuncttrue
\mciteSetBstMidEndSepPunct{\mcitedefaultmidpunct}
{\mcitedefaultendpunct}{\mcitedefaultseppunct}\relax
\EndOfBibitem
\bibitem{ATLAS:2017wvi}
ATLAS collaboration, M.~Aaboud {\em et~al.}, \ifthenelse{\boolean{articletitles}}{\emph{{Measurement of the inclusive and fiducial $t\bar{t}$ production cross-sections in the lepton+jets channel in $pp$ collisions at $\sqrt{s} = 8$ TeV with the ATLAS detector}}, }{}\href{https://doi.org/10.1140/epjc/s10052-018-5904-z}{Eur.\ Phys.\ J.\  \textbf{C78} (2018) 487}, \href{http://arxiv.org/abs/1712.06857}{{\normalfont\ttfamily arXiv:1712.06857}}\relax
\mciteBstWouldAddEndPuncttrue
\mciteSetBstMidEndSepPunct{\mcitedefaultmidpunct}
{\mcitedefaultendpunct}{\mcitedefaultseppunct}\relax
\EndOfBibitem
\bibitem{ATLAS:2017rso}
ATLAS collaboration, M.~Aaboud {\em et~al.}, \ifthenelse{\boolean{articletitles}}{\emph{{Fiducial, total and differential cross-section measurements of $t$-channel single top-quark production in $pp$ collisions at 8 TeV using data collected by the ATLAS detector}}, }{}\href{https://doi.org/10.1140/epjc/s10052-017-5061-9}{Eur.\ Phys.\ J.\  \textbf{C77} (2017) 531}, \href{http://arxiv.org/abs/1702.02859}{{\normalfont\ttfamily arXiv:1702.02859}}\relax
\mciteBstWouldAddEndPuncttrue
\mciteSetBstMidEndSepPunct{\mcitedefaultmidpunct}
{\mcitedefaultendpunct}{\mcitedefaultseppunct}\relax
\EndOfBibitem
\bibitem{CMS:2018fks}
CMS collaboration, A.~M. Sirunyan {\em et~al.}, \ifthenelse{\boolean{articletitles}}{\emph{{Measurement of the $t\bar{t}$ production cross section, the top quark mass, and the strong coupling constant using dilepton events in pp collisions at $\sqrt{s} =$ 13 TeV}}, }{}\href{https://doi.org/10.1140/epjc/s10052-019-6863-8}{Eur.\ Phys.\ J.\  \textbf{C79} (2019) 368}, \href{http://arxiv.org/abs/1812.10505}{{\normalfont\ttfamily arXiv:1812.10505}}\relax
\mciteBstWouldAddEndPuncttrue
\mciteSetBstMidEndSepPunct{\mcitedefaultmidpunct}
{\mcitedefaultendpunct}{\mcitedefaultseppunct}\relax
\EndOfBibitem
\bibitem{CMS:2018lgn}
CMS collaboration, A.~M. Sirunyan {\em et~al.}, \ifthenelse{\boolean{articletitles}}{\emph{{Measurement of the single top quark and antiquark production cross sections in the $t$ channel and their ratio in proton-proton collisions at $\sqrt{s}=$ 13 TeV}}, }{}\href{https://doi.org/10.1016/j.physletb.2019.135042}{Phys.\ Lett.\  \textbf{B800} (2020) 135042}, \href{http://arxiv.org/abs/1812.10514}{{\normalfont\ttfamily arXiv:1812.10514}}\relax
\mciteBstWouldAddEndPuncttrue
\mciteSetBstMidEndSepPunct{\mcitedefaultmidpunct}
{\mcitedefaultendpunct}{\mcitedefaultseppunct}\relax
\EndOfBibitem
\bibitem{CMS:2019snc}
CMS collaboration, A.~M. Sirunyan {\em et~al.}, \ifthenelse{\boolean{articletitles}}{\emph{{Measurement of the top quark pair production cross section in dilepton final states containing one $\tau$ lepton in pp collisions at $\sqrt{s}=$ 13 TeV}}, }{}\href{https://doi.org/10.1007/JHEP02(2020)191}{JHEP \textbf{02} (2020) 191}, \href{http://arxiv.org/abs/1911.13204}{{\normalfont\ttfamily arXiv:1911.13204}}\relax
\mciteBstWouldAddEndPuncttrue
\mciteSetBstMidEndSepPunct{\mcitedefaultmidpunct}
{\mcitedefaultendpunct}{\mcitedefaultseppunct}\relax
\EndOfBibitem
\bibitem{ATLAS:2020ccu}
ATLAS collaboration, G.~Aad {\em et~al.}, \ifthenelse{\boolean{articletitles}}{\emph{{Measurements of top-quark pair single- and double-differential cross-sections in the all-hadronic channel in $pp$ collisions at $\sqrt{s}=13~\textrm{TeV}$ using the ATLAS detector}}, }{}\href{https://doi.org/10.1007/JHEP01(2021)033}{JHEP \textbf{01} (2021) 033}, \href{http://arxiv.org/abs/2006.09274}{{\normalfont\ttfamily arXiv:2006.09274}}\relax
\mciteBstWouldAddEndPuncttrue
\mciteSetBstMidEndSepPunct{\mcitedefaultmidpunct}
{\mcitedefaultendpunct}{\mcitedefaultseppunct}\relax
\EndOfBibitem
\bibitem{CMS:2021gwv}
CMS collaboration, A.~Tumasyan {\em et~al.}, \ifthenelse{\boolean{articletitles}}{\emph{{Measurement of the inclusive $t\bar{t}$ production cross section in proton-proton collisions at $ \sqrt{s} $ = 5.02 TeV}}, }{}\href{https://doi.org/10.1007/JHEP04(2022)144}{JHEP \textbf{04} (2022) 144}, \href{http://arxiv.org/abs/2112.09114}{{\normalfont\ttfamily arXiv:2112.09114}}\relax
\mciteBstWouldAddEndPuncttrue
\mciteSetBstMidEndSepPunct{\mcitedefaultmidpunct}
{\mcitedefaultendpunct}{\mcitedefaultseppunct}\relax
\EndOfBibitem
\bibitem{CMS:2021vhb}
CMS collaboration, A.~Tumasyan {\em et~al.}, \ifthenelse{\boolean{articletitles}}{\emph{{Measurement of differential $t\bar{t}$ production cross sections in the full kinematic range using lepton+jets events from proton-proton collisions at $\sqrt {s}$ = 13\,\,TeV}}, }{}\href{https://doi.org/10.1103/PhysRevD.104.092013}{Phys.\ Rev.\  \textbf{D104} (2021) 092013}, \href{http://arxiv.org/abs/2108.02803}{{\normalfont\ttfamily arXiv:2108.02803}}\relax
\mciteBstWouldAddEndPuncttrue
\mciteSetBstMidEndSepPunct{\mcitedefaultmidpunct}
{\mcitedefaultendpunct}{\mcitedefaultseppunct}\relax
\EndOfBibitem
\bibitem{ATLAS:2022aof}
ATLAS, CMS collaborations, G.~Aad {\em et~al.}, \ifthenelse{\boolean{articletitles}}{\emph{{Combination of inclusive top-quark pair production cross-section measurements using ATLAS and CMS data at $ \sqrt{s} $ = 7 and 8 TeV}}, }{}\href{https://doi.org/10.1007/JHEP07(2023)213}{JHEP \textbf{07} (2023) 213}, \href{http://arxiv.org/abs/2205.13830}{{\normalfont\ttfamily arXiv:2205.13830}}\relax
\mciteBstWouldAddEndPuncttrue
\mciteSetBstMidEndSepPunct{\mcitedefaultmidpunct}
{\mcitedefaultendpunct}{\mcitedefaultseppunct}\relax
\EndOfBibitem
\bibitem{ATLAS:2022jbj}
ATLAS collaboration, G.~Aad {\em et~al.}, \ifthenelse{\boolean{articletitles}}{\emph{{Measurement of the $ t\overline{t} $ production cross-section in pp collisions at $ \sqrt{s} $ = 5.02 TeV with the ATLAS detector}}, }{}\href{https://doi.org/10.1007/JHEP06(2023)138}{JHEP \textbf{06} (2023) 138}, \href{http://arxiv.org/abs/2207.01354}{{\normalfont\ttfamily arXiv:2207.01354}}\relax
\mciteBstWouldAddEndPuncttrue
\mciteSetBstMidEndSepPunct{\mcitedefaultmidpunct}
{\mcitedefaultendpunct}{\mcitedefaultseppunct}\relax
\EndOfBibitem
\bibitem{ATLAS:2022opp}
ATLAS collaboration, M.~Aaboud {\em et~al.}, \ifthenelse{\boolean{articletitles}}{\emph{{Measurement of the inclusive $t\bar{t}$ production cross section in the lepton+jets channel in pp collisions at $\sqrt{s} =$7{\,}{\,}TeV with the ATLAS detector using support vector machines}}, }{}\href{https://doi.org/10.1103/PhysRevD.108.032014}{Phys.\ Rev.\  \textbf{D108} (2023) 032014}, \href{http://arxiv.org/abs/2212.00571}{{\normalfont\ttfamily arXiv:2212.00571}}\relax
\mciteBstWouldAddEndPuncttrue
\mciteSetBstMidEndSepPunct{\mcitedefaultmidpunct}
{\mcitedefaultendpunct}{\mcitedefaultseppunct}\relax
\EndOfBibitem
\bibitem{ATLAS:2023gsl}
ATLAS collaboration, G.~Aad {\em et~al.}, \ifthenelse{\boolean{articletitles}}{\emph{{Inclusive and differential cross-sections for dilepton $ t\overline{t} $ production measured in $ \sqrt{s} $ = 13 TeV pp collisions with the ATLAS detector}}, }{}\href{https://doi.org/10.1007/JHEP07(2023)141}{JHEP \textbf{07} (2023) 141}, \href{http://arxiv.org/abs/2303.15340}{{\normalfont\ttfamily arXiv:2303.15340}}\relax
\mciteBstWouldAddEndPuncttrue
\mciteSetBstMidEndSepPunct{\mcitedefaultmidpunct}
{\mcitedefaultendpunct}{\mcitedefaultseppunct}\relax
\EndOfBibitem
\bibitem{ATLAS:2023slx}
ATLAS collaboration, G.~Aad {\em et~al.}, \ifthenelse{\boolean{articletitles}}{\emph{{Measurement of the $t\bar{t}$ cross section and its ratio to the $Z$ production cross section using $pp$ collisions at $\sqrt{s} =$13.6 TeV with the ATLAS detector}}, }{}\href{https://doi.org/10.1016/j.physletb.2023.138376}{Phys.\ Lett.\  \textbf{B848} (2024) 138376}, \href{http://arxiv.org/abs/2308.09529}{{\normalfont\ttfamily arXiv:2308.09529}}\relax
\mciteBstWouldAddEndPuncttrue
\mciteSetBstMidEndSepPunct{\mcitedefaultmidpunct}
{\mcitedefaultendpunct}{\mcitedefaultseppunct}\relax
\EndOfBibitem
\bibitem{CMS:2023qyl}
CMS collaboration, A.~Tumasyan {\em et~al.}, \ifthenelse{\boolean{articletitles}}{\emph{{First measurement of the top quark pair production cross section in proton-proton collisions at $ \sqrt{s} $ = 13.6 TeV}}, }{}\href{https://doi.org/10.1007/JHEP08(2023)204}{JHEP \textbf{08} (2023) 204}, \href{http://arxiv.org/abs/2303.10680}{{\normalfont\ttfamily arXiv:2303.10680}}\relax
\mciteBstWouldAddEndPuncttrue
\mciteSetBstMidEndSepPunct{\mcitedefaultmidpunct}
{\mcitedefaultendpunct}{\mcitedefaultseppunct}\relax
\EndOfBibitem
\bibitem{CMS:2024ghc}
CMS collaboration, A.~Hayrapetyan {\em et~al.}, \ifthenelse{\boolean{articletitles}}{\emph{{Measurement of the inclusive $t\bar{t}$ cross section in final states with at least one lepton and additional jets with 302 pb$^{-1}$ of pp collisions at $\sqrt{s}$ = 5.02 TeV}}, }{}\href{https://doi.org/10.1007/JHEP04(2025)099}{JHEP \textbf{04} (2025) 099}, \href{http://arxiv.org/abs/2410.21631}{{\normalfont\ttfamily arXiv:2410.21631}}\relax
\mciteBstWouldAddEndPuncttrue
\mciteSetBstMidEndSepPunct{\mcitedefaultmidpunct}
{\mcitedefaultendpunct}{\mcitedefaultseppunct}\relax
\EndOfBibitem
\bibitem{LHCb-PAPER-2015-022}
LHCb collaboration, R.~Aaij {\em et~al.}, \ifthenelse{\boolean{articletitles}}{\emph{{First observation of top quark production in the forward region}}, }{}\href{https://doi.org/10.1103/PhysRevLett.115.112001}{Phys.\ Rev.\ Lett.\  \textbf{115} (2015) 112001}, \href{http://arxiv.org/abs/1506.00903}{{\normalfont\ttfamily arXiv:1506.00903}}\relax
\mciteBstWouldAddEndPuncttrue
\mciteSetBstMidEndSepPunct{\mcitedefaultmidpunct}
{\mcitedefaultendpunct}{\mcitedefaultseppunct}\relax
\EndOfBibitem
\bibitem{LHCb-PAPER-2016-038}
LHCb collaboration, R.~Aaij {\em et~al.}, \ifthenelse{\boolean{articletitles}}{\emph{{Measurement of forward $\ttbar$, $W + \bbbar$ and $W + \ccbar$ production in \proton\proton collisions at \mbox{$\sqs=$8~\tev}}}, }{}\href{https://doi.org/10.1016/j.physletb.2017.01.044}{Phys.\ Lett.\  \textbf{B767} (2017) 110}, \href{http://arxiv.org/abs/1610.08142}{{\normalfont\ttfamily arXiv:1610.08142}}\relax
\mciteBstWouldAddEndPuncttrue
\mciteSetBstMidEndSepPunct{\mcitedefaultmidpunct}
{\mcitedefaultendpunct}{\mcitedefaultseppunct}\relax
\EndOfBibitem
\bibitem{LHCb-PAPER-2017-050}
LHCb collaboration, R.~Aaij {\em et~al.}, \ifthenelse{\boolean{articletitles}}{\emph{{Measurement of forward top pair production in the dilepton channel in \proton\proton collisions at $\sqs=13~\tev$}}, }{}\href{https://doi.org/10.1007/JHEP08(2018)174}{JHEP \textbf{08} (2018) 174}, \href{http://arxiv.org/abs/1803.05188}{{\normalfont\ttfamily arXiv:1803.05188}}\relax
\mciteBstWouldAddEndPuncttrue
\mciteSetBstMidEndSepPunct{\mcitedefaultmidpunct}
{\mcitedefaultendpunct}{\mcitedefaultseppunct}\relax
\EndOfBibitem
\bibitem{CDF:2011xdt}
CDF collaboration, T.~Aaltonen {\em et~al.}, \ifthenelse{\boolean{articletitles}}{\emph{{Evidence for a mass dependent forward-backward asymmetry in top quark pair production}}, }{}\href{https://doi.org/10.1103/PhysRevD.83.112003}{Phys.\ Rev.\  \textbf{D83} (2011) 112003}, \href{http://arxiv.org/abs/1101.0034}{{\normalfont\ttfamily arXiv:1101.0034}}\relax
\mciteBstWouldAddEndPuncttrue
\mciteSetBstMidEndSepPunct{\mcitedefaultmidpunct}
{\mcitedefaultendpunct}{\mcitedefaultseppunct}\relax
\EndOfBibitem
\bibitem{D0:2015alu}
D0 collaboration, V.~M. Abazov {\em et~al.}, \ifthenelse{\boolean{articletitles}}{\emph{{Simultaneous measurement of forward-backward asymmetry and top polarization in dilepton final states from $t\bar t$ production at the Tevatron}}, }{}\href{https://doi.org/10.1103/PhysRevD.92.052007}{Phys.\ Rev.\  \textbf{D92} (2015) 052007}, \href{http://arxiv.org/abs/1507.05666}{{\normalfont\ttfamily arXiv:1507.05666}}\relax
\mciteBstWouldAddEndPuncttrue
\mciteSetBstMidEndSepPunct{\mcitedefaultmidpunct}
{\mcitedefaultendpunct}{\mcitedefaultseppunct}\relax
\EndOfBibitem
\bibitem{CDF:2012ctl}
CDF collaboration, T.~Aaltonen {\em et~al.}, \ifthenelse{\boolean{articletitles}}{\emph{{Measurement of the top quark forward-backward production asymmetry and its dependence on event kinematic properties}}, }{}\href{https://doi.org/10.1103/PhysRevD.87.092002}{Phys.\ Rev.\  \textbf{D87} (2013) 092002}, \href{http://arxiv.org/abs/1211.1003}{{\normalfont\ttfamily arXiv:1211.1003}}\relax
\mciteBstWouldAddEndPuncttrue
\mciteSetBstMidEndSepPunct{\mcitedefaultmidpunct}
{\mcitedefaultendpunct}{\mcitedefaultseppunct}\relax
\EndOfBibitem
\bibitem{Czakon:2014xsa}
M.~Czakon, P.~Fiedler, and A.~Mitov, \ifthenelse{\boolean{articletitles}}{\emph{{Resolving the Tevatron top quark forward-backward asymmetry puzzle: fully differential next-to-next-to-leading-order calculation}}, }{}\href{https://doi.org/10.1103/PhysRevLett.115.052001}{Phys.\ Rev.\ Lett.\  \textbf{115} (2015) 052001}, \href{http://arxiv.org/abs/1411.3007}{{\normalfont\ttfamily arXiv:1411.3007}}\relax
\mciteBstWouldAddEndPuncttrue
\mciteSetBstMidEndSepPunct{\mcitedefaultmidpunct}
{\mcitedefaultendpunct}{\mcitedefaultseppunct}\relax
\EndOfBibitem
\bibitem{D0:2014cda}
D0 collaboration, V.~M. Abazov {\em et~al.}, \ifthenelse{\boolean{articletitles}}{\emph{{Measurement of the forward-backward asymmetry in top quark-antiquark production in $p\bar{p}$ collisions using the lepton+jets channel}}, }{}\href{https://doi.org/10.1103/PhysRevD.90.072011}{Phys.\ Rev.\  \textbf{D90} (2014) 072011}, \href{http://arxiv.org/abs/1405.0421}{{\normalfont\ttfamily arXiv:1405.0421}}\relax
\mciteBstWouldAddEndPuncttrue
\mciteSetBstMidEndSepPunct{\mcitedefaultmidpunct}
{\mcitedefaultendpunct}{\mcitedefaultseppunct}\relax
\EndOfBibitem
\bibitem{Kidonakis:2015ona}
N.~Kidonakis, \ifthenelse{\boolean{articletitles}}{\emph{{Top quark forward-backward asymmetry at approximate N$^3$LO}}, }{}\href{https://doi.org/10.1103/PhysRevD.91.071502}{Phys.\ Rev.\  \textbf{D91} (2015) 071502(R)}, \href{http://arxiv.org/abs/1501.01581}{{\normalfont\ttfamily arXiv:1501.01581}}\relax
\mciteBstWouldAddEndPuncttrue
\mciteSetBstMidEndSepPunct{\mcitedefaultmidpunct}
{\mcitedefaultendpunct}{\mcitedefaultseppunct}\relax
\EndOfBibitem
\bibitem{CDF:2017cvy}
CDF, D0 collaborations, T.~A. Aaltonen {\em et~al.}, \ifthenelse{\boolean{articletitles}}{\emph{{Combined forward-backward asymmetry measurements in top-antitop quark production at the Tevatron}}, }{}\href{https://doi.org/10.1103/PhysRevLett.120.042001}{Phys.\ Rev.\ Lett.\  \textbf{120} (2018) 042001}, \href{http://arxiv.org/abs/1709.04894}{{\normalfont\ttfamily arXiv:1709.04894}}\relax
\mciteBstWouldAddEndPuncttrue
\mciteSetBstMidEndSepPunct{\mcitedefaultmidpunct}
{\mcitedefaultendpunct}{\mcitedefaultseppunct}\relax
\EndOfBibitem
\bibitem{ATLAS:2013buu}
ATLAS collaboration, G.~Aad {\em et~al.}, \ifthenelse{\boolean{articletitles}}{\emph{{Measurement of the top quark pair production charge asymmetry in proton-proton collisions at $\sqrt{s}$ = 7 TeV using the ATLAS detector}}, }{}\href{https://doi.org/10.1007/JHEP02(2014)107}{JHEP \textbf{02} (2014) 107}, \href{http://arxiv.org/abs/1311.6724}{{\normalfont\ttfamily arXiv:1311.6724}}\relax
\mciteBstWouldAddEndPuncttrue
\mciteSetBstMidEndSepPunct{\mcitedefaultmidpunct}
{\mcitedefaultendpunct}{\mcitedefaultseppunct}\relax
\EndOfBibitem
\bibitem{ATLAS:2015ysm}
ATLAS collaboration, G.~Aad {\em et~al.}, \ifthenelse{\boolean{articletitles}}{\emph{{Measurement of the charge asymmetry in dileptonic decays of top quark pairs in $pp$ collisions at $\sqrt{s}=7$ TeV using the ATLAS detector}}, }{}\href{https://doi.org/10.1007/JHEP05(2015)061}{JHEP \textbf{05} (2015) 061}, \href{http://arxiv.org/abs/1501.07383}{{\normalfont\ttfamily arXiv:1501.07383}}\relax
\mciteBstWouldAddEndPuncttrue
\mciteSetBstMidEndSepPunct{\mcitedefaultmidpunct}
{\mcitedefaultendpunct}{\mcitedefaultseppunct}\relax
\EndOfBibitem
\bibitem{ATLAS:2015jgj}
ATLAS collaboration, G.~Aad {\em et~al.}, \ifthenelse{\boolean{articletitles}}{\emph{{Measurement of the charge asymmetry in top-quark pair production in the lepton-plus-jets final state in pp collision data at $\sqrt{s}=8\,\mathrm TeV{}$ with the ATLAS detector}}, }{}\href{https://doi.org/10.1140/epjc/s10052-016-3910-6}{Eur.\ Phys.\ J.\  \textbf{C76} (2016) 87}, Erratum \href{https://doi.org/10.1140/epjc/s10052-017-5089-x}{ibid.\   \textbf{C77} (2017) 564}, \href{http://arxiv.org/abs/1509.02358}{{\normalfont\ttfamily arXiv:1509.02358}}\relax
\mciteBstWouldAddEndPuncttrue
\mciteSetBstMidEndSepPunct{\mcitedefaultmidpunct}
{\mcitedefaultendpunct}{\mcitedefaultseppunct}\relax
\EndOfBibitem
\bibitem{ATLAS:2015sex}
ATLAS collaboration, G.~Aad {\em et~al.}, \ifthenelse{\boolean{articletitles}}{\emph{{Measurement of the charge asymmetry in highly boosted top-quark pair production in $\sqrt{s} =$ 8 TeV $pp$ collision data collected by the ATLAS experiment}}, }{}\href{https://doi.org/10.1016/j.physletb.2016.02.055}{Phys.\ Lett.\  \textbf{B756} (2016) 52}, \href{http://arxiv.org/abs/1512.06092}{{\normalfont\ttfamily arXiv:1512.06092}}\relax
\mciteBstWouldAddEndPuncttrue
\mciteSetBstMidEndSepPunct{\mcitedefaultmidpunct}
{\mcitedefaultendpunct}{\mcitedefaultseppunct}\relax
\EndOfBibitem
\bibitem{ATLAS:2022wec}
ATLAS collaboration, G.~Aad {\em et~al.}, \ifthenelse{\boolean{articletitles}}{\emph{{Measurement of the charge asymmetry in top-quark pair production in association with a photon with the ATLAS experiment}}, }{}\href{https://doi.org/10.1016/j.physletb.2023.137848}{Phys.\ Lett.\  \textbf{B843} (2023) 137848}, \href{http://arxiv.org/abs/2212.10552}{{\normalfont\ttfamily arXiv:2212.10552}}\relax
\mciteBstWouldAddEndPuncttrue
\mciteSetBstMidEndSepPunct{\mcitedefaultmidpunct}
{\mcitedefaultendpunct}{\mcitedefaultseppunct}\relax
\EndOfBibitem
\bibitem{ATLAS:2022waa}
ATLAS collaboration, G.~Aad {\em et~al.}, \ifthenelse{\boolean{articletitles}}{\emph{{Evidence for the charge asymmetry in pp {\textrightarrow} $ t\overline{t} $ production at $ \sqrt{s} $ = 13 TeV with the ATLAS detector}}, }{}\href{https://doi.org/10.1007/JHEP08(2023)077}{JHEP \textbf{08} (2023) 077}, \href{http://arxiv.org/abs/2208.12095}{{\normalfont\ttfamily arXiv:2208.12095}}\relax
\mciteBstWouldAddEndPuncttrue
\mciteSetBstMidEndSepPunct{\mcitedefaultmidpunct}
{\mcitedefaultendpunct}{\mcitedefaultseppunct}\relax
\EndOfBibitem
\bibitem{CMS:2012oht}
CMS collaboration, S.~Chatrchyan {\em et~al.}, \ifthenelse{\boolean{articletitles}}{\emph{{Inclusive and differential measurements of the $t \bar{t}$ charge asymmetry in proton-proton collisions at $\sqrt{s} =$ 7 TeV}}, }{}\href{https://doi.org/10.1016/j.physletb.2012.09.028}{Phys.\ Lett.\  \textbf{B717} (2012) 129}, \href{http://arxiv.org/abs/1207.0065}{{\normalfont\ttfamily arXiv:1207.0065}}\relax
\mciteBstWouldAddEndPuncttrue
\mciteSetBstMidEndSepPunct{\mcitedefaultmidpunct}
{\mcitedefaultendpunct}{\mcitedefaultseppunct}\relax
\EndOfBibitem
\bibitem{CMS:2014rdf}
CMS collaboration, S.~Chatrchyan {\em et~al.}, \ifthenelse{\boolean{articletitles}}{\emph{{Measurements of the $t\bar{t}$ charge asymmetry using the dilepton decay channel in pp collisions at $\sqrt{s} =$ 7 TeV}}, }{}\href{https://doi.org/10.1007/JHEP04(2014)191}{JHEP \textbf{04} (2014) 191}, \href{http://arxiv.org/abs/1402.3803}{{\normalfont\ttfamily arXiv:1402.3803}}\relax
\mciteBstWouldAddEndPuncttrue
\mciteSetBstMidEndSepPunct{\mcitedefaultmidpunct}
{\mcitedefaultendpunct}{\mcitedefaultseppunct}\relax
\EndOfBibitem
\bibitem{CMS:2015fvy}
CMS collaboration, V.~Khachatryan {\em et~al.}, \ifthenelse{\boolean{articletitles}}{\emph{{Measurement of the charge asymmetry in top quark pair production in pp collisions at $\sqrt{s} =$ 8 TeV using a template method}}, }{}\href{https://doi.org/10.1103/PhysRevD.93.034014}{Phys.\ Rev.\  \textbf{D93} (2016) 034014}, \href{http://arxiv.org/abs/1508.03862}{{\normalfont\ttfamily arXiv:1508.03862}}\relax
\mciteBstWouldAddEndPuncttrue
\mciteSetBstMidEndSepPunct{\mcitedefaultmidpunct}
{\mcitedefaultendpunct}{\mcitedefaultseppunct}\relax
\EndOfBibitem
\bibitem{CMS:2015pob}
CMS collaboration, V.~Khachatryan {\em et~al.}, \ifthenelse{\boolean{articletitles}}{\emph{{Inclusive and differential measurements of the $t\bar{t}$ charge asymmetry in pp collisions at $\sqrt{s} =$ 8 TeV}}, }{}\href{https://doi.org/10.1016/j.physletb.2016.03.060}{Phys.\ Lett.\  \textbf{B757} (2016) 154}, \href{http://arxiv.org/abs/1507.03119}{{\normalfont\ttfamily arXiv:1507.03119}}\relax
\mciteBstWouldAddEndPuncttrue
\mciteSetBstMidEndSepPunct{\mcitedefaultmidpunct}
{\mcitedefaultendpunct}{\mcitedefaultseppunct}\relax
\EndOfBibitem
\bibitem{CMS:2016ypc}
CMS collaboration, V.~Khachatryan {\em et~al.}, \ifthenelse{\boolean{articletitles}}{\emph{{Measurements of $t \bar t$ charge asymmetry using dilepton final states in pp collisions at $\sqrt s=8$ TeV}}, }{}\href{https://doi.org/10.1016/j.physletb.2016.07.006}{Phys.\ Lett.\  \textbf{B760} (2016) 365}, \href{http://arxiv.org/abs/1603.06221}{{\normalfont\ttfamily arXiv:1603.06221}}\relax
\mciteBstWouldAddEndPuncttrue
\mciteSetBstMidEndSepPunct{\mcitedefaultmidpunct}
{\mcitedefaultendpunct}{\mcitedefaultseppunct}\relax
\EndOfBibitem
\bibitem{antiKt}
M.~Cacciari, G.~P. Salam, and G.~Soyez, \ifthenelse{\boolean{articletitles}}{\emph{{The anti-$k_t$ jet clustering algorithm}}, }{}\href{https://doi.org/10.1088/1126-6708/2008/04/063}{JHEP \textbf{04} (2008) 063}, \href{http://arxiv.org/abs/0802.1189}{{\normalfont\ttfamily arXiv:0802.1189}}\relax
\mciteBstWouldAddEndPuncttrue
\mciteSetBstMidEndSepPunct{\mcitedefaultmidpunct}
{\mcitedefaultendpunct}{\mcitedefaultseppunct}\relax
\EndOfBibitem
\bibitem{LHCb-PAPER-2013-058}
LHCb collaboration, R.~Aaij {\em et~al.}, \ifthenelse{\boolean{articletitles}}{\emph{{Study of forward \Z+jet production in \proton\proton collisions at $\sqs = $7~\tev}}, }{}\href{https://doi.org/10.1007/JHEP01(2014)033}{JHEP \textbf{01} (2014) 033}, \href{http://arxiv.org/abs/1310.8197}{{\normalfont\ttfamily arXiv:1310.8197}}\relax
\mciteBstWouldAddEndPuncttrue
\mciteSetBstMidEndSepPunct{\mcitedefaultmidpunct}
{\mcitedefaultendpunct}{\mcitedefaultseppunct}\relax
\EndOfBibitem
\bibitem{LHCb-DP-2008-001}
LHCb collaboration, A.~A. Alves~Jr.\ {\em et~al.}, \ifthenelse{\boolean{articletitles}}{\emph{{The \lhcb detector at the LHC}}, }{}\href{https://doi.org/10.1088/1748-0221/3/08/S08005}{JINST \textbf{3} (2008) S08005}\relax
\mciteBstWouldAddEndPuncttrue
\mciteSetBstMidEndSepPunct{\mcitedefaultmidpunct}
{\mcitedefaultendpunct}{\mcitedefaultseppunct}\relax
\EndOfBibitem
\bibitem{LHCb-DP-2014-002}
LHCb collaboration, R.~Aaij {\em et~al.}, \ifthenelse{\boolean{articletitles}}{\emph{{LHCb detector performance}}, }{}\href{https://doi.org/10.1142/S0217751X15300227}{Int.\ J.\ Mod.\ Phys.\  \textbf{A30} (2015) 1530022}, \href{http://arxiv.org/abs/1412.6352}{{\normalfont\ttfamily arXiv:1412.6352}}\relax
\mciteBstWouldAddEndPuncttrue
\mciteSetBstMidEndSepPunct{\mcitedefaultmidpunct}
{\mcitedefaultendpunct}{\mcitedefaultseppunct}\relax
\EndOfBibitem
\bibitem{LHCb-DP-2014-001}
R.~Aaij {\em et~al.}, \ifthenelse{\boolean{articletitles}}{\emph{{Performance of the LHCb Vertex Locator}}, }{}\href{https://doi.org/10.1088/1748-0221/9/09/P09007}{JINST \textbf{9} (2014) P09007}, \href{http://arxiv.org/abs/1405.7808}{{\normalfont\ttfamily arXiv:1405.7808}}\relax
\mciteBstWouldAddEndPuncttrue
\mciteSetBstMidEndSepPunct{\mcitedefaultmidpunct}
{\mcitedefaultendpunct}{\mcitedefaultseppunct}\relax
\EndOfBibitem
\bibitem{LHCb-DP-2017-001}
P.~d'Argent {\em et~al.}, \ifthenelse{\boolean{articletitles}}{\emph{{Improved performance of the LHCb Outer Tracker in LHC Run 2}}, }{}\href{https://doi.org/10.1088/1748-0221/12/11/P11016}{JINST \textbf{12} (2017) P11016}, \href{http://arxiv.org/abs/1708.00819}{{\normalfont\ttfamily arXiv:1708.00819}}\relax
\mciteBstWouldAddEndPuncttrue
\mciteSetBstMidEndSepPunct{\mcitedefaultmidpunct}
{\mcitedefaultendpunct}{\mcitedefaultseppunct}\relax
\EndOfBibitem
\bibitem{LHCb-DP-2012-003}
M.~Adinolfi {\em et~al.}, \ifthenelse{\boolean{articletitles}}{\emph{{Performance of the \lhcb RICH detector at the LHC}}, }{}\href{https://doi.org/10.1140/epjc/s10052-013-2431-9}{Eur.\ Phys.\ J.\  \textbf{C73} (2013) 2431}, \href{http://arxiv.org/abs/1211.6759}{{\normalfont\ttfamily arXiv:1211.6759}}\relax
\mciteBstWouldAddEndPuncttrue
\mciteSetBstMidEndSepPunct{\mcitedefaultmidpunct}
{\mcitedefaultendpunct}{\mcitedefaultseppunct}\relax
\EndOfBibitem
\bibitem{LHCb-DP-2012-002}
A.~A. Alves~Jr.\ {\em et~al.}, \ifthenelse{\boolean{articletitles}}{\emph{{Performance of the LHCb muon system}}, }{}\href{https://doi.org/10.1088/1748-0221/8/02/P02022}{JINST \textbf{8} (2013) P02022}, \href{http://arxiv.org/abs/1211.1346}{{\normalfont\ttfamily arXiv:1211.1346}}\relax
\mciteBstWouldAddEndPuncttrue
\mciteSetBstMidEndSepPunct{\mcitedefaultmidpunct}
{\mcitedefaultendpunct}{\mcitedefaultseppunct}\relax
\EndOfBibitem
\bibitem{LHCb-DP-2012-004}
R.~Aaij {\em et~al.}, \ifthenelse{\boolean{articletitles}}{\emph{{The \lhcb trigger and its performance in 2011}}, }{}\href{https://doi.org/10.1088/1748-0221/8/04/P04022}{JINST \textbf{8} (2013) P04022}, \href{http://arxiv.org/abs/1211.3055}{{\normalfont\ttfamily arXiv:1211.3055}}\relax
\mciteBstWouldAddEndPuncttrue
\mciteSetBstMidEndSepPunct{\mcitedefaultmidpunct}
{\mcitedefaultendpunct}{\mcitedefaultseppunct}\relax
\EndOfBibitem
\bibitem{LHCb-DP-2019-001}
R.~Aaij {\em et~al.}, \ifthenelse{\boolean{articletitles}}{\emph{{Design and performance of the LHCb trigger and full real-time reconstruction in Run 2 of the LHC}}, }{}\href{https://doi.org/10.1088/1748-0221/14/04/P04013}{JINST \textbf{14} (2019) P04013}, \href{http://arxiv.org/abs/1812.10790}{{\normalfont\ttfamily arXiv:1812.10790}}\relax
\mciteBstWouldAddEndPuncttrue
\mciteSetBstMidEndSepPunct{\mcitedefaultmidpunct}
{\mcitedefaultendpunct}{\mcitedefaultseppunct}\relax
\EndOfBibitem
\bibitem{Sjostrand:2007gs}
T.~Sj\"{o}strand, S.~Mrenna, and P.~Skands, \ifthenelse{\boolean{articletitles}}{\emph{{A brief introduction to PYTHIA 8.1}}, }{}\href{https://doi.org/10.1016/j.cpc.2008.01.036}{Comput.\ Phys.\ Commun.\  \textbf{178} (2008) 852}, \href{http://arxiv.org/abs/0710.3820}{{\normalfont\ttfamily arXiv:0710.3820}}\relax
\mciteBstWouldAddEndPuncttrue
\mciteSetBstMidEndSepPunct{\mcitedefaultmidpunct}
{\mcitedefaultendpunct}{\mcitedefaultseppunct}\relax
\EndOfBibitem
\bibitem{Sjostrand:2006za}
T.~Sj\"{o}strand, S.~Mrenna, and P.~Skands, \ifthenelse{\boolean{articletitles}}{\emph{{PYTHIA 6.4 physics and manual}}, }{}\href{https://doi.org/10.1088/1126-6708/2006/05/026}{JHEP \textbf{05} (2006) 026}, \href{http://arxiv.org/abs/hep-ph/0603175}{{\normalfont\ttfamily arXiv:hep-ph/0603175}}\relax
\mciteBstWouldAddEndPuncttrue
\mciteSetBstMidEndSepPunct{\mcitedefaultmidpunct}
{\mcitedefaultendpunct}{\mcitedefaultseppunct}\relax
\EndOfBibitem
\bibitem{LHCb-PROC-2010-056}
I.~Belyaev {\em et~al.}, \ifthenelse{\boolean{articletitles}}{\emph{{Handling of the generation of primary events in Gauss, the LHCb simulation framework}}, }{}\href{https://doi.org/10.1088/1742-6596/331/3/032047}{J.\ Phys.\ Conf.\ Ser.\  \textbf{331} (2011) 032047}\relax
\mciteBstWouldAddEndPuncttrue
\mciteSetBstMidEndSepPunct{\mcitedefaultmidpunct}
{\mcitedefaultendpunct}{\mcitedefaultseppunct}\relax
\EndOfBibitem
\bibitem{davidson2015photos}
N.~Davidson, T.~Przedzinski, and Z.~Was, \ifthenelse{\boolean{articletitles}}{\emph{{PHOTOS interface in C++: Technical and physics documentation}}, }{}\href{https://doi.org/https://doi.org/10.1016/j.cpc.2015.09.013}{Comp.\ Phys.\ Comm.\  \textbf{199} (2016) 86}, \href{http://arxiv.org/abs/1011.0937}{{\normalfont\ttfamily arXiv:1011.0937}}\relax
\mciteBstWouldAddEndPuncttrue
\mciteSetBstMidEndSepPunct{\mcitedefaultmidpunct}
{\mcitedefaultendpunct}{\mcitedefaultseppunct}\relax
\EndOfBibitem
\bibitem{Allison:2006ve}
Geant4 collaboration, J.~Allison {\em et~al.}, \ifthenelse{\boolean{articletitles}}{\emph{{Geant4 developments and applications}}, }{}\href{https://doi.org/10.1109/TNS.2006.869826}{IEEE Trans.\ Nucl.\ Sci.\  \textbf{53} (2006) 270}\relax
\mciteBstWouldAddEndPuncttrue
\mciteSetBstMidEndSepPunct{\mcitedefaultmidpunct}
{\mcitedefaultendpunct}{\mcitedefaultseppunct}\relax
\EndOfBibitem
\bibitem{Agostinelli:2002hh}
Geant4 collaboration, S.~Agostinelli {\em et~al.}, \ifthenelse{\boolean{articletitles}}{\emph{{Geant4: A simulation toolkit}}, }{}\href{https://doi.org/10.1016/S0168-9002(03)01368-8}{Nucl.\ Instrum.\ Meth.\  \textbf{A506} (2003) 250}\relax
\mciteBstWouldAddEndPuncttrue
\mciteSetBstMidEndSepPunct{\mcitedefaultmidpunct}
{\mcitedefaultendpunct}{\mcitedefaultseppunct}\relax
\EndOfBibitem
\bibitem{LHCb-PROC-2011-006}
M.~Clemencic {\em et~al.}, \ifthenelse{\boolean{articletitles}}{\emph{{The \lhcb simulation application, Gauss: Design, evolution and experience}}, }{}\href{https://doi.org/10.1088/1742-6596/331/3/032023}{J.\ Phys.\ Conf.\ Ser.\  \textbf{331} (2011) 032023}\relax
\mciteBstWouldAddEndPuncttrue
\mciteSetBstMidEndSepPunct{\mcitedefaultmidpunct}
{\mcitedefaultendpunct}{\mcitedefaultseppunct}\relax
\EndOfBibitem
\bibitem{Nason:2004rx}
P.~Nason, \ifthenelse{\boolean{articletitles}}{\emph{{A new method for combining NLO QCD with shower Monte Carlo algorithms}}, }{}\href{https://doi.org/10.1088/1126-6708/2004/11/040}{JHEP \textbf{11} (2004) 040}, \href{http://arxiv.org/abs/hep-ph/0409146}{{\normalfont\ttfamily arXiv:hep-ph/0409146}}\relax
\mciteBstWouldAddEndPuncttrue
\mciteSetBstMidEndSepPunct{\mcitedefaultmidpunct}
{\mcitedefaultendpunct}{\mcitedefaultseppunct}\relax
\EndOfBibitem
\bibitem{Frixione:2007vw}
S.~Frixione, P.~Nason, and C.~Oleari, \ifthenelse{\boolean{articletitles}}{\emph{{Matching NLO QCD computations with parton shower simulations: the POWHEG method}}, }{}\href{https://doi.org/10.1088/1126-6708/2007/11/070}{JHEP \textbf{11} (2007) 070}, \href{http://arxiv.org/abs/0709.2092}{{\normalfont\ttfamily arXiv:0709.2092}}\relax
\mciteBstWouldAddEndPuncttrue
\mciteSetBstMidEndSepPunct{\mcitedefaultmidpunct}
{\mcitedefaultendpunct}{\mcitedefaultseppunct}\relax
\EndOfBibitem
\bibitem{Alioli:2010xd}
S.~Alioli, P.~Nason, C.~Oleari, and E.~Re, \ifthenelse{\boolean{articletitles}}{\emph{{A general framework for implementing NLO calculations in shower Monte Carlo programs: the POWHEG BOX}}, }{}\href{https://doi.org/10.1007/JHEP06(2010)043}{JHEP \textbf{06} (2010) 043}, \href{http://arxiv.org/abs/1002.2581}{{\normalfont\ttfamily arXiv:1002.2581}}\relax
\mciteBstWouldAddEndPuncttrue
\mciteSetBstMidEndSepPunct{\mcitedefaultmidpunct}
{\mcitedefaultendpunct}{\mcitedefaultseppunct}\relax
\EndOfBibitem
\bibitem{Mg5}
J.~Alwall {\em et~al.}, \ifthenelse{\boolean{articletitles}}{\emph{{The automated computation of tree-level and next-to-leading order differential cross sections, and their matching to parton shower simulations}}, }{}\href{https://doi.org/10.1007/JHEP07(2014)079}{JHEP \textbf{07} (2014) 079}, \href{http://arxiv.org/abs/1405.0301}{{\normalfont\ttfamily arXiv:1405.0301}}\relax
\mciteBstWouldAddEndPuncttrue
\mciteSetBstMidEndSepPunct{\mcitedefaultmidpunct}
{\mcitedefaultendpunct}{\mcitedefaultseppunct}\relax
\EndOfBibitem
\bibitem{LHCb-PAPER-2016-011}
LHCb collaboration, R.~Aaij {\em et~al.}, \ifthenelse{\boolean{articletitles}}{\emph{{Measurement of forward $\W$ and $\Z$ boson production in association with jets in proton-proton collisions at \mbox{$\sqs=$8~\tev}}}, }{}\href{https://doi.org/10.1007/JHEP05(2016)131}{JHEP \textbf{05} (2016) 131}, \href{http://arxiv.org/abs/1605.00951}{{\normalfont\ttfamily arXiv:1605.00951}}\relax
\mciteBstWouldAddEndPuncttrue
\mciteSetBstMidEndSepPunct{\mcitedefaultmidpunct}
{\mcitedefaultendpunct}{\mcitedefaultseppunct}\relax
\EndOfBibitem
\bibitem{deepjet}
E.~Bols {\em et~al.}, \ifthenelse{\boolean{articletitles}}{\emph{{Jet flavour classification using DeepJet}}, }{}\href{https://doi.org/10.1088/1748-0221/15/12/P12012}{JINST \textbf{15} (2020) P12012}, \href{http://arxiv.org/abs/2008.10519}{{\normalfont\ttfamily arXiv:2008.10519}}\relax
\mciteBstWouldAddEndPuncttrue
\mciteSetBstMidEndSepPunct{\mcitedefaultmidpunct}
{\mcitedefaultendpunct}{\mcitedefaultseppunct}\relax
\EndOfBibitem
\bibitem{LHCb-PAPER-2025-034}
LHCb collaboration, R.~Aaij {\em et~al.}, \ifthenelse{\boolean{articletitles}}{\emph{{Search for $H \to b \Bar{b}$ and $H \to c \Bar{c}$ in $\sqrt{s}=13$ TeV $pp$ collisions at LHCb using machine learning techniques}}, }{} {LHCb-PAPER-2025-034}, {in preparation, to be submitted to JHEP}\relax
\mciteBstWouldAddEndPuncttrue
\mciteSetBstMidEndSepPunct{\mcitedefaultmidpunct}
{\mcitedefaultendpunct}{\mcitedefaultseppunct}\relax
\EndOfBibitem
\bibitem{ABCD_ATLAS}
ATLAS collaboration, G.~Aad {\em et~al.}, \ifthenelse{\boolean{articletitles}}{\emph{{Measurement of the inclusive isolated prompt photon cross section in $pp$ collisions at $\sqrt{s}=7$ TeV with the ATLAS detector}}, }{}\href{https://doi.org/10.1103/PhysRevD.83.052005}{Phys.\ Rev.\  \textbf{D83} (2011) 052005}, \href{http://arxiv.org/abs/1012.4389}{{\normalfont\ttfamily arXiv:1012.4389}}\relax
\mciteBstWouldAddEndPuncttrue
\mciteSetBstMidEndSepPunct{\mcitedefaultmidpunct}
{\mcitedefaultendpunct}{\mcitedefaultseppunct}\relax
\EndOfBibitem
\bibitem{LHCB-PAPER-2021-037}
LHCb collaboration, R.~Aaij {\em et~al.}, \ifthenelse{\boolean{articletitles}}{\emph{{Precision measurement of forward \Z boson production in proton-proton collisions at $\sqrt{s} = 13$~\tev}}, }{}\href{https://doi.org/10.1007/JHEP07(2022)026}{JHEP \textbf{07} (2022) 026}, \href{http://arxiv.org/abs/2112.07458}{{\normalfont\ttfamily arXiv:2112.07458}}\relax
\mciteBstWouldAddEndPuncttrue
\mciteSetBstMidEndSepPunct{\mcitedefaultmidpunct}
{\mcitedefaultendpunct}{\mcitedefaultseppunct}\relax
\EndOfBibitem
\bibitem{LHCB-PAPER-2014-033}
LHCb collaboration, R.~Aaij {\em et~al.}, \ifthenelse{\boolean{articletitles}}{\emph{{Measurement of the forward \W boson production cross-section in \proton\proton collisions at \mbox{$\sqs=$7~\tev}}}, }{}\href{https://doi.org/10.1007/JHEP12(2014)079}{JHEP \textbf{12} (2014) 079}, \href{http://arxiv.org/abs/1408.4354}{{\normalfont\ttfamily arXiv:1408.4354}}\relax
\mciteBstWouldAddEndPuncttrue
\mciteSetBstMidEndSepPunct{\mcitedefaultmidpunct}
{\mcitedefaultendpunct}{\mcitedefaultseppunct}\relax
\EndOfBibitem
\bibitem{LHCB-PAPER-2015-001}
LHCb collaboration, R.~Aaij {\em et~al.}, \ifthenelse{\boolean{articletitles}}{\emph{{Measurement of the forward \Z boson cross-section in \proton\proton collisions at \mbox{$\sqs=$7~\tev}}}, }{}\href{https://doi.org/10.1007/JHEP08(2015)039}{JHEP \textbf{08} (2015) 039}, \href{http://arxiv.org/abs/1505.07024}{{\normalfont\ttfamily arXiv:1505.07024}}\relax
\mciteBstWouldAddEndPuncttrue
\mciteSetBstMidEndSepPunct{\mcitedefaultmidpunct}
{\mcitedefaultendpunct}{\mcitedefaultseppunct}\relax
\EndOfBibitem
\bibitem{LHCB-PAPER-2015-049}
LHCb collaboration, R.~Aaij {\em et~al.}, \ifthenelse{\boolean{articletitles}}{\emph{{Measurement of forward \W and \Z boson production in \proton\proton collisions at \mbox{$\sqs=$8~\tev}}}, }{}\href{https://doi.org/10.1007/JHEP01(2016)155}{JHEP \textbf{01} (2016) 155}, \href{http://arxiv.org/abs/1511.08039}{{\normalfont\ttfamily arXiv:1511.08039}}\relax
\mciteBstWouldAddEndPuncttrue
\mciteSetBstMidEndSepPunct{\mcitedefaultmidpunct}
{\mcitedefaultendpunct}{\mcitedefaultseppunct}\relax
\EndOfBibitem
\bibitem{LHCb-PAPER-2020-018}
LHCb collaboration, R.~Aaij {\em et~al.}, \ifthenelse{\boolean{articletitles}}{\emph{{Measurement of differential $\bquark \bquarkbar$ and $\cquark \cquarkbar$ dijet cross-sections in the forward region of $pp$ collisions at $\sqrt{s} = 13$~\tev}}, }{}\href{https://doi.org/10.1007/JHEP02(2021)023}{JHEP \textbf{02} (2021) 023}, \href{http://arxiv.org/abs/2010.09437}{{\normalfont\ttfamily arXiv:2010.09437}}\relax
\mciteBstWouldAddEndPuncttrue
\mciteSetBstMidEndSepPunct{\mcitedefaultmidpunct}
{\mcitedefaultendpunct}{\mcitedefaultseppunct}\relax
\EndOfBibitem
\bibitem{LHCb-PAPER-2015-021}
LHCb collaboration, R.~Aaij {\em et~al.}, \ifthenelse{\boolean{articletitles}}{\emph{{Study of \W boson production in association with beauty and charm}}, }{}\href{https://doi.org/10.1103/PhysRevD.92.052001}{Phys.\ Rev.\  \textbf{D92} (2015) 052012}, \href{http://arxiv.org/abs/1505.04051}{{\normalfont\ttfamily arXiv:1505.04051}}\relax
\mciteBstWouldAddEndPuncttrue
\mciteSetBstMidEndSepPunct{\mcitedefaultmidpunct}
{\mcitedefaultendpunct}{\mcitedefaultseppunct}\relax
\EndOfBibitem
\bibitem{theoryerr}
LHC Higgs Cross Section Working Group, S.~Dittmaier {\em et~al.}, \ifthenelse{\boolean{articletitles}}{\emph{{Handbook of LHC Higgs cross sections: 1. inclusive observables}}, }{}doi:~\href{https://doi.org/10.5170/CERN-2011-002}{10.5170/CERN-2011-002} \href{http://arxiv.org/abs/1101.0593}{{\normalfont\ttfamily arXiv:1101.0593}}\relax
\mciteBstWouldAddEndPuncttrue
\mciteSetBstMidEndSepPunct{\mcitedefaultmidpunct}
{\mcitedefaultendpunct}{\mcitedefaultseppunct}\relax
\EndOfBibitem
\bibitem{lumi}
LHCb collaboration, R.~Aaij {\em et~al.}, \ifthenelse{\boolean{articletitles}}{\emph{{Precision luminosity measurements at LHCb}}, }{}\href{https://doi.org/10.1088/1748-0221/9/12/P12005}{JINST \textbf{9} (2014) P12005}, \href{http://arxiv.org/abs/1410.0149}{{\normalfont\ttfamily arXiv:1410.0149}}\relax
\mciteBstWouldAddEndPuncttrue
\mciteSetBstMidEndSepPunct{\mcitedefaultmidpunct}
{\mcitedefaultendpunct}{\mcitedefaultseppunct}\relax
\EndOfBibitem
\bibitem{POWHEG}
T.~Je\v{z}o and P.~Nason, \ifthenelse{\boolean{articletitles}}{\emph{{On the treatment of resonances in next-to-leading order calculations matched to a parton shower}}, }{}\href{https://doi.org/10.1007/JHEP12(2015)065}{JHEP \textbf{12} (2015) 065}, \href{http://arxiv.org/abs/1509.09071}{{\normalfont\ttfamily arXiv:1509.09071}}\relax
\mciteBstWouldAddEndPuncttrue
\mciteSetBstMidEndSepPunct{\mcitedefaultmidpunct}
{\mcitedefaultendpunct}{\mcitedefaultseppunct}\relax
\EndOfBibitem
\bibitem{Hou:2019efy}
T.-J. Hou {\em et~al.}, \ifthenelse{\boolean{articletitles}}{\emph{{New CTEQ global analysis of quantum chromodynamics with high-precision data from the LHC}}, }{}\href{https://doi.org/10.1103/PhysRevD.103.014013}{Phys.\ Rev.\  \textbf{D103} (2021) 014013}, \href{http://arxiv.org/abs/1912.10053}{{\normalfont\ttfamily arXiv:1912.10053}}\relax
\mciteBstWouldAddEndPuncttrue
\mciteSetBstMidEndSepPunct{\mcitedefaultmidpunct}
{\mcitedefaultendpunct}{\mcitedefaultseppunct}\relax
\EndOfBibitem
\bibitem{Ball2017}
NNPDF collaboration, R.~D. Ball {\em et~al.}, \ifthenelse{\boolean{articletitles}}{\emph{{Parton distributions from high-precision collider data}}, }{}\href{https://doi.org/10.1140/epjc/s10052-017-5199-5}{Eur.\ Phys.\ J.\  \textbf{C77} (2017) 663}, \href{http://arxiv.org/abs/1706.00428}{{\normalfont\ttfamily arXiv:1706.00428}}\relax
\mciteBstWouldAddEndPuncttrue
\mciteSetBstMidEndSepPunct{\mcitedefaultmidpunct}
{\mcitedefaultendpunct}{\mcitedefaultseppunct}\relax
\EndOfBibitem
\bibitem{bourilkov2006}
D.~Bourilkov, R.~C. Group, and M.~R. Whalley, \ifthenelse{\boolean{articletitles}}{\emph{{LHAPDF: PDF use from the Tevatron to the LHC}}, }{} in {\em {TeV4LHC Workshop - 4th meeting}}, 2006, \href{http://arxiv.org/abs/hep-ph/0605240}{{\normalfont\ttfamily arXiv:hep-ph/0605240}}\relax
\mciteBstWouldAddEndPuncttrue
\mciteSetBstMidEndSepPunct{\mcitedefaultmidpunct}
{\mcitedefaultendpunct}{\mcitedefaultseppunct}\relax
\EndOfBibitem
\bibitem{LHCb-PAPER-2025-057-cds}
\lhcb Collaboration, \ifthenelse{\boolean{articletitles}}{\emph{{Measurement of the top-quark production cross-section and charge asymmetry at LHCb}}, }{} \href{https://cds.cern.ch/record/2951095} {LHCb-PAPER-2025-057-cds}\relax
\mciteBstWouldAddEndPuncttrue
\mciteSetBstMidEndSepPunct{\mcitedefaultmidpunct}
{\mcitedefaultendpunct}{\mcitedefaultseppunct}\relax
\EndOfBibitem
\bibitem{james_paper}
J.~V. Mead, \ifthenelse{\boolean{articletitles}}{\emph{{A measurement of top quark production at $ \sqrt{s} $ = 13 TeV with LHCb data}}, }{} 2021.
\newblock \href{https://cds.cern.ch/record/2780780} {CERN-THESIS-2021-130}\relax
\mciteBstWouldAddEndPuncttrue
\mciteSetBstMidEndSepPunct{\mcitedefaultmidpunct}
{\mcitedefaultendpunct}{\mcitedefaultseppunct}\relax
\EndOfBibitem
\end{mcitethebibliography}
\ifx\mcitethebibliography\mciteundefinedmacro
\PackageError{LHCb.bst}{mciteplus.sty has not been loaded}
{This bibstyle requires the use of the mciteplus package.}\fi
\providecommand{\href}[2]{#2}

\clearpage
\centerline
{\large\bf LHCb collaboration}
\begin
{flushleft}
\small
R.~Aaij$^{38}$\lhcborcid{0000-0003-0533-1952},
A.S.W.~Abdelmotteleb$^{58}$\lhcborcid{0000-0001-7905-0542},
C.~Abellan~Beteta$^{52}$\lhcborcid{0009-0009-0869-6798},
F.~Abudin{\'e}n$^{60}$\lhcborcid{0000-0002-6737-3528},
T.~Ackernley$^{62}$\lhcborcid{0000-0002-5951-3498},
A. A. ~Adefisoye$^{70}$\lhcborcid{0000-0003-2448-1550},
B.~Adeva$^{48}$\lhcborcid{0000-0001-9756-3712},
M.~Adinolfi$^{56}$\lhcborcid{0000-0002-1326-1264},
P.~Adlarson$^{86}$\lhcborcid{0000-0001-6280-3851},
C.~Agapopoulou$^{14}$\lhcborcid{0000-0002-2368-0147}
% ...

{\footnotesize \it

% ...
$^{14}$Universit{\'e} Paris-Saclay, CNRS/IN2P3, IJCLab, Orsay, France\\
% ...
$^{38}$Nikhef National Institute for Subatomic Physics, Amsterdam, Netherlands\\
% ...
$^{48}$Instituto Galego de F{\'\i}sica de Altas Enerx{\'\i}as (IGFAE), Universidade de Santiago de Compostela, Santiago de Compostela, Spain\\
% ...
$^{52}$Physik-Institut, Universit{\"a}t Z{\"u}rich, Z{\"u}rich, Switzerland\\
% ...
$^{56}$H.H. Wills Physics Laboratory, University of Bristol, Bristol, United Kingdom\\
% ...
$^{58}$Department of Physics, University of Warwick, Coventry, United Kingdom\\
% ...
$^{60}$School of Physics and Astronomy, University of Edinburgh, Edinburgh, United Kingdom\\
$^{61}$School of Physics and Astronomy, University of Glasgow, Glasgow, United Kingdom\\
$^{62}$Oliver Lodge Laboratory, University of Liverpool, Liverpool, United Kingdom\\
% ...
$^{70}$Syracuse University, Syracuse, NY, United States\\
% ...
$^{86}$Department of Physics and Astronomy, Uppsala University, Uppsala, Sweden, associated to $^{61}$\\
% ...
\medskip
}
\end{flushleft}





\end{document}